\documentclass[10pt,notitlepage,nofootinbib]{revtex4-1}

\usepackage[latin1]{inputenc}
\usepackage[T1]{fontenc}
\usepackage{amsmath,amsfonts,amssymb,amsthm,url,geometry,mathrsfs,color,esint,bm}
\usepackage{mathtools}
\usepackage{enumerate}
\usepackage[pdftex]{hyperref}
\usepackage{dsfont}
\usepackage{mathdots}
\usepackage{mathtools}
\usepackage{stackengine}
\usepackage{scalerel}
\usepackage{graphicx}
\usepackage{array}
\usepackage{tikz}
\usepackage{makecell}
\newcolumntype{x}[1]{>{\centering\arraybackslash}p{#1}}
\usepackage{braket}
\usepackage{chngcntr}
\usepackage{apptools}
\usepackage{titlesec}
\usepackage[detect-all]{siunitx}

\titleformat{\section}{\bfseries\boldmath\centering}{\thesection.}{1em}{\MakeUppercase}

\newcommand{\e}{\varepsilon}
\newcommand{\iy}{\infty}
\newcommand{\st}{\ : \ }
\newcommand{\id}{\mathrm{Id}}

\newcommand\op[1]{\widetilde{#1}}

\newcommand{\bb}{\begin{equation}}
\newcommand{\ee}{\end{equation}}
\newcommand{\bbb}{\begin{equation*}}
\newcommand{\eee}{\end{equation*}}

\renewcommand{\leq}{\leqslant}
\renewcommand{\geq}{\geqslant}

\newcommand{\gM}{\mathsf{M}}

\newcommand*{\coloneqq}{\mathrel{\vcenter{\baselineskip0.5ex \lineskiplimit0pt \hbox{\scriptsize.}\hbox{\scriptsize.}}} =}
\newcommand{\texteq}[1]{\stackrel{\mathclap{\scriptsize \mbox{#1}}}{=}}
\newcommand{\textleq}[1]{\stackrel{\mathclap{\scriptsize \mbox{#1}}}{\leq}}

\newcommand{\sumno}{\sum\nolimits}
\newcommand{\ketbra}[2]{\ket{#1}\!\!\bra{#2}}

\DeclareMathOperator{\Tr}{tr}

\DeclareMathOperator{\tr}{tr}
\DeclareMathOperator{\rk}{rk}
\DeclareMathOperator{\cl}{cl}
\DeclareMathOperator{\dbm}{d}
\DeclareMathOperator{\f}{f}
\DeclareMathOperator{\wf}{wf}
\DeclareMathOperator{\weakd}{wd}
\DeclareMathAlphabet{\pazocal}{OMS}{zplm}{m}{n}

\newcommand{\tmin}{\ensuremath \! \raisebox{2.6pt}{$\underset{\begin{array}{c} \vspace{-3.7ex} \\ \text{\scriptsize min} \end{array}}{\otimes}$}\!}
\newcommand{\tmax}{\ensuremath \! \raisebox{2.6pt}{$\underset{\begin{array}{c} \vspace{-3.7ex} \\ \text{\scriptsize max} \end{array}}{\otimes}$}\! }
\newcommand{\tminit}{\ensuremath \! \raisebox{2.6pt}{$\underset{\begin{array}{c} \vspace{-3.7ex} \\ \text{\scriptsize \emph{min}} \end{array}}{\otimes}$}\!}

\newcommand{\tminfoot}{\ensuremath \!\! \raisebox{2.0pt}{$\scriptstyle\underset{\begin{array}{c} \vspace{-4.0ex} \\ \text{\fontsize{2}{4}\selectfont min} \end{array}}{\otimes}$}\!\!}

\newcommand{\tmaxfoot}{\ensuremath \!\! \raisebox{2.0pt}{$\scriptstyle\underset{\begin{array}{c} \vspace{-4.0ex} \\ \text{\fontsize{2}{4}\selectfont max} \end{array}}{\otimes}$}\!\!}

\renewcommand{\P}{\mathbf{P}}
\newcommand{\R}{\mathbf{R}}

\newcommand{\A}{\mathcal{A}}
\newcommand{\B}{\mathcal{B}}

\DeclareMathOperator{\E}{\mathbf{E}}
\DeclareMathOperator{\conv}{\mathrm{conv}}

\DeclareMathOperator{\mathspan}{\mathrm{span}}

\newcommand{\scalar}[2]{\langle #1 , #2\rangle}

\newcommand{\Tn}{S_1^{n,\mathrm{sa}}}
\newcommand{\Tm}{S_1^{m,\mathrm{sa}}}
\newcommand{\Tk}{S_1^{k,\mathrm{sa}}}
\newcommand{\Herm}[1]{\gM_{#1}^{\mathrm{sa}}}

\AtAppendix{\counterwithin{theorem}{section}}

\theoremstyle{plain}
\newtheorem{theorem}{Theorem}
\newtheorem{proposition}[theorem]{Proposition}
\newtheorem{lemma}[theorem]{Lemma}

\newtheorem{problem}[theorem]{Problem}
\newtheorem{corollary}[theorem]{Corollary}
\newtheorem{definition}{Definition}

\theoremstyle{definition}
\newtheorem*{remark}{Remark}
\newtheorem*{note}{Note}

\begin{document}

%\title{Fantastic tensor norms and where to find them}
\title{Universal gaps for XOR games from estimates on tensor norm ratios}

\author{Guillaume Aubrun}
\email{aubrun@math.univ-lyon1.fr}
\affiliation{Institut Camille Jordan, Universit\'e Claude Bernard Lyon 1, 43 boulevard du 11 novembre 1918, 69622 Villeurbanne CEDEX, France}

\author{Ludovico Lami}
\email{ludovico.lami@gmail.com}
\affiliation{School of Mathematical Sciences and Centre for the Mathematics and Theoretical Physics of Quantum Non-Equilibrium Systems, University of Nottingham, University Park, Nottingham NG7 2RD, United Kingdom}

\author{Carlos Palazuelos}
\email{cpalazue@ucm.es}
\affiliation{Departamento de An\'alisis Matem\'atico y Matem\'atica Aplicada, Universidad Complutense de Madrid, Plaza de Ciencias
s/n 28040 Madrid, Spain,}
\affiliation{Instituto de Ciencias Matem\'aticas, C/ Nicol\'as Cabrera, 13-15, 28049
Madrid, Spain}

\author{Stanis\l aw J. Szarek}
\email{szarek@cwru.edu}
\affiliation{Case Western Reserve University, Department of Mathematics, Applied Mathematics and Statistics, 10900 Euclid Avenue, Cleveland, Ohio 44106,  USA}
\affiliation{Sorbonne Universit\'e, Institut de Math\'ematiques de Jussieu-PRG, 4 place Jussieu, 75005 Paris, France}

\author{Andreas Winter}
\email{andreas.winter@uab.cat}
\affiliation{F\'isica Te\`orica: Informaci\'o i Fen\`omens Qu\`antics, Departament de F\'isica, Universitat Aut\`onoma de Barcelona, ES-08193 Bellaterra (Barcelona), Spain}
\affiliation{ICREA -- Instituci\'o Catalana de Recerca i Estudis Avan\c{c}ats, Pg.\ Lluis Companys 23, ES-08010 Barcelona, Spain}

\begin{abstract}
    We define and study XOR games in the framework of general probabilistic theories, which encompasses all physical models whose predictive power obeys minimal requirements. The bias of an XOR game under local or global strategies is shown to be given by a certain injective or projective tensor norm, respectively. The intrinsic (i.e.\ model-independent) advantage of global over local strategies is thus connected to a universal function $r(n,m)$ called `projective-injective ratio'. This is defined as the minimal constant $\rho$ such that $\|\cdot\|_{X\otimes_\pi Y}\leq\rho\,\|\cdot\|_{X\otimes_\varepsilon Y}$ holds for all Banach spaces of dimensions $\dim X=n$ and $\dim Y=m$, where $X\otimes_\pi Y$ and $X \otimes_\varepsilon Y$ are the projective and injective tensor products. By requiring that $X=Y$, one obtains a symmetrised version of the above ratio, denoted by $r_s(n)$. We prove that $r(n,m)\geq 19/18$ for all $n,m\geq 2$, implying that injective and projective tensor products are never isometric. We then study the asymptotic behaviour of $r(n,m)$ and $r_s(n)$, showing that, up to log factors: $r_s(n)$ is of the order $\sqrt{n}$ (which is sharp); $r(n,n)$ is at least of the order $n^{1/6}$; and $r(n,m)$ grows at least as $\min\{n,m\}^{1/8}$. These results constitute our main contribution to the theory of tensor norms. In our proof, a crucial role is played by an `$\ell_1$/$\ell_2$/$\ell_{\iy}$ trichotomy theorem' based on ideas by Pisier, Rudelson, Szarek, and Tomczak-Jaegermann. The main operational consequence we draw is that there is a universal gap between local and global strategies in general XOR games, and that this grows as a power of the minimal local dimension. In the quantum case, we are able to determine this gap up to universal constants. As a corollary, we obtain an improved bound on the scaling of the maximal quantum data hiding efficiency against local measurements.
\end{abstract}

\maketitle

%\tableofcontents

\section{Introduction}

%https://www.overleaf.com/1346819983cffwxcrmsfcn
%\begin{enumerate}[(i)]
%\item General motivation.
%\item General probabilistic theories: basic + composites + base norms: all indispensable definitions and no more.
%\item Tensor norms, projective and injective. Projective = maximal GPT base norm.
%\item XOR games: classical and quantum games as special case of GPT games. Formula for bias = injective norm. Relation with state discrimination. Mention bounded communication. Main question (qualitative, for $X\otimes X$).
%\item Pi/Epsilon ratio: mathematical motivation and definitions ($\rho,R,r,r_s$). $R$ can be computed exactly.
%\item Main results: universal constants ($\Rightarrow$ no isometry for Pisier's space), asymptotics for $r_s$ and $r$. Solution of $S_1^n\otimes S_1^m$.
%\item Structure of the paper.
%\end{enumerate}

One of the most prominent conceptual contributions of the celebrated 1964 paper by J.S. Bell~\cite{Bell} is to point out that the implications of the quantum mechanical predictions extend far beyond the very same formalism that is used to deduce them, and shed light on some of the deepest secrets of Nature. 
As is well known, the scenario considered by Bell features two distant parties that share a quantum entangled state and make local quantum measurements, but the conclusions of the argument, far from depending on the details of the involved physics, rely only upon the correctness of its experimental predictions, which are expressed in terms of entirely classical input-output relations.
%As is well known, the scenario considered by Bell features two distant parties that share a quantum entangled state, but the conclusions of the argument, far from depending on the details of the involved physics, rely only upon the correctness of its experimental predictions.
In the same spirit, we deem it important to understand what features of information processing in composite systems are truly \textit{intrinsic}, meaning that they are common to all conceivable physical theories.

A suitable way to formalise the concept of a physical theory in this context is provided by the mathematical machinery of general probabilistic theories (GPTs)~\cite{FOUNDATIONS, Barrett-original, lamiatesi}. It is sometimes convenient to think of GPTs as generalisations of finite-al quantum mechanics, where the set of unnormalised states is not assumed to be the cone of positive semidefinite matrices, but it is rather taken to be an arbitrary convex cone in a finite-al real vector space. As is well known, a GPT makes the host vector space a Banach space in a canonical way, by equipping it with a so-called base norm.
%Through an appropriate extension of the Born rule, it is possible to employ this formalism to make probabilistic predictions on the outcome of experiments on physical systems. We provide a brief introduction to the formalism below, while referring to~\cite[Section~2]{ultimate} for a more detailed presentation.

The starting point of our investigation is the study of XOR games in the rich landscape of GPTs. We remind the reader that any XOR game can be equivalently cast in terms of state discrimination queries subjected to locality constraints, so that our analysis applies equally well to these problems.
XOR games are arguably the simplest examples of two-prover one-round games and feature two players, Alice and Bob, and a third party known as the referee. The referee asks the players some `questions' by sending them states of some physical system modelled by a GPT. The correctness of the one-bit answers the players provide upon measuring the state depends only on their parity. According to whether Alice and Bob are allowed to carry out product or global measurements, one talks about local or global strategies. In general, the winning probability can be significantly larger in the latter than in the former case.
%Although Alice and Bob are forbidden to communicate after the questions have been asked, they may be allowed to share a bipartite physical system, prepared in some appropriate state as modelled by a GPT, and use the information retrieved from it via local measurements to help deciding the answers. Despite the correlations extracted from the shared state have to obey the no-signalling principle, they can be used to effectively increase the winning probability.

While the quantitative details of this phenomenon will in general depend on the particular physical system modelling the questions, our work is instead motivated by the wish to understand which behaviours are universal, and thus pertain to the intrinsic nature of XOR games. This line of investigation brings us to develop an extensive connection with the theory of tensor norms on finite-dimensional Banach spaces, which has already proved to be instrumental in the study of classical and quantum XOR games~\cite{PV16}. While in these more standard settings one deals with \textit{specific examples} of tensor norms, the analysis of games played over arbitrary GPT models requires a \textit{systematic} understanding of \textit{general} tensor norms.
%We are thus naturally led to identify some deep questions in the general theory, which we then proceed to investigate in the second part of the paper.

The main problem we investigate here asks for the maximal gap that can be guaranteed to exist between the winning probabilities associated with global and local strategies in XOR games played over GPTs of fixed local s. In analogy with the classical case, we show that such winning probabilities are given by simple expressions involving respectively the projective and injective tensor norms induced by the local GPTs through their native Banach space structures. Comparing them in a model-independent fashion prompts us to investigate the least constant of domination of the injective over the projective tensor norm over all pairs of normed spaces of fixed s. We call this function \textit{`projective/injective ratio'}, or \textit{`$\pi/\e$ ratio'} for short. When seen from the point of view of pure mathematics, this universal function over pairs of positive integers encodes some information regarding Grothendieck's theory of tensor products of Banach spaces. At the same time, the operational interpretation we construct here guarantees that the same object captures some intrinsic feature of general XOR games.

%We prove a number of results about the projective/injective ratio and related quantities.
Our main result is that the $\pi/\e$ ratio associated with two $n$-al Banach spaces scales at least as $n^{1/6}$ (up to logarithmic factors), implying that global strategies for XOR games are intrinsically much more effective than local ones in a precise asymptotic sense. We ask the question whether this scaling can be improved up to $n^{1/2}$, and bolster this hypothesis by showing that it holds true (again, up to log factors) when two copies of the same space are considered. Interestingly, this question is intimately connected to the problem of estimating the radius of the weak Banach--Mazur compactum, which has also been conjectured to be of order $n^{1/2}$~\cite{Tomczak84, Rudelson95}.
We also consider the problem of establishing universal lower bounds for the $\pi/\e$ ratio. We prove that for all pairs of Banach spaces $X$, $Y$ there is a tensor in $X\otimes Y$ whose projective norm is at least $19/18$ times its injective norm. In particular, these norms are always different. This should be compared with the famous construction by Pisier~\cite{Pisier} of an infinite-al Banach space $X$ such that the injective and projective norms on $X \otimes X$ are equivalent.
Finally, we solve the problem of computing the $\pi/\e$ ratio for some specific examples of physically relevant Banach spaces. Most notably, we establish that it is of the order $\min\{n,m\}^{3/2}$ for $\Tn \otimes \Tm$, where $\Tk$ stands for the space of $k\times k$ Hermitian matrices endowed with the trace norm. The importance of this special case stems from the fact that $\Tk$ is the natural Banach space associated with a $k$-level quantum system.

The rest of the paper is structured as follows. Throughout this section, we provide very brief introductions to the GPT formalism (Subsection~\ref{subsec GPT}), to the theory of tensor norms (Subsection~\ref{subsec tensor norms}), and to XOR games (Subsection~\ref{subsec XOR}). In Section~\ref{sec main results}, we state our main results and broadly discuss some of the proof techniques we developed. Section~\ref{sec first bounds} presents some general properties of the $\pi/\e$ ratio, connecting it with other concepts in functional analysis. There, we find the universal lower bound of $19/18$, and solve the quantum mechanical case up to multiplicative constants. Section~\ref{sec asymptotic bounds} deals with the problem of determining the asymptotic scaling of the $\pi/\e$ ratio, either for two copies of the same space, or in the fully general setting of two normed spaces of different s. Extra information can be found in Appendices. Appendix~\ref{app XOR games in GPTs} investigates how our operational interpretation of the injective tensor norm is affected by the introduction of a bounded amount of two-way communication, while Appendix~\ref{app base norm equivalence} provides a proof of the useful fact that any normed space is $2$-isomorphic to a base norm space. Finally, Appendix~\ref{app functional analysis} gathers the functional-analytic background that is used throughout the paper and which may be unfamiliar to a non-specialist reader.

\subsection{General probabilistic theories} \label{subsec GPT}

The origins of the formalism of general probabilistic theories (GPTs) lie in the attempt to axiomatise quantum mechanics, rebuilding it upon operationally motivated postulates rather than upon more evasive concepts such as `wave function' and `microscopic system'. Although these ideas can be found already in some antecedent works~\cite{Segal1947, MACKEY}, the first major contributions were made by the `Marburg school' led by Ludwig~\cite{Ludwig-1, LUDWIG}. This resulted in an intense debate around the nascent GPT formalism, which took place in a series of papers published in Communications in Mathematical Physics~\cite{Ludwig-2, Ludwig-3, Daehn-4, Stolz-5, Davies-1970, Edwards-operational}. For an account of the early development of the field, we refer the interested reader to~\cite{FOUNDATIONS}. A more modern point of view can be found in~\cite[Chapter~1]{lamiatesi}.
%The debate around the nascent GPT formalism took place in a series of papers published in Communications in Mathematical Physics~\cite{Ludwig-2, Ludwig-3, Daehn-4, Stolz-5, Mielnik-filters, Davies-1970, Edwards-operational, Stolz-6, Edwards-operations, Edwards-simple-observables, Ludwig-7, Edwards-pure-operations, Daehn-algebra-filters, Daehn-symm-modularity, Gudder-convex-structure, Mielnik-general-quantum}, for an account of which we refer the reader to~\cite{FOUNDATIONS}.
This foundationally motivated interest has seen a revival in the last two decades, with much effort being focused on attempts to `reconstruct' quantum mechanics starting from first principles~\cite{Hardy2001, DAriano-1, Wilce-4-1/2-axioms, Masanes2011, no-3-order}. At the same time, GPTs have become central to quantum information science, as they provide indispensable tools to analyse information processing beyond classical theories, see for instance~\cite{PR-boxes, nonloc-resource, nonloc-comm-complex, PVV, Barnum-no-broad, Barnum-steering, telep-in-GPT, Jencova2017, ultimate}.
%\cite{PR-boxes, nonloc-resource, nonloc-comm-complex, PVV, Barrett-original, Barnum-no-broad, Barnum-steering, telep-in-GPT, Jencova2017, ultimate}.
An introduction to the GPT framework that will suffice for our purposes can be found in~\cite[Chapter~2]{lamiatesi} (see also~\cite[Section~2]{ultimate}). Throughout this subsection, we limit ourselves to recalling the basics and to fixing the notation.

\begin{definition}
A \textbf{general probabilistic theory} is a triple $(V,C,u)$, where: (i) $V$ is a finite-al real vector space; (ii) $C\subset V$ is a closed, convex, salient and generating cone; and (iii) $u$, called the \textbf{order unit} or the \textbf{unit effect}, is a functional in the interior of the dual cone $C^*\coloneqq \{x^* \in V^*:\, x^*(x)\geq 0\ \forall\, x\in C \}$. GPTs will be denoted by capital letters such as $A$, $B$ etc., which -- with a slight abuse of notation -- identify also the underlying physical systems. We call $\dim V$ the \textbf{} of the GPT.
\end{definition}

On the mathematical level, we can think of $(V,C)$ as an ordered vector space, the ordering being given by $x\leq y\Leftrightarrow y-x\in C$. Also the dual vector space can be thought of as ordered by the dual cone $C^*$. In this language, the functional $u$ is said to be strictly positive, since $x\geq 0$ and $u(x)=0$ implies $x=0$. The states of the physical system modelled by $(V,C,u)$ are represented by vectors in $C\cap u^{-1}(1)\eqqcolon \Omega$. The compact convex set $\Omega$ is called the \textbf{state space} of the GPT, and accordingly we will sometimes refer to $C$ as the cone of unnormalised states. Convexity here plays an operationally relevant role, as the process of preparing a system in a state $\omega_0$ with probability $p$ and $\omega_1$ with probability $1-p$, and later forgetting the value of the binary random variable associated with its preparation, leaves the system in the state $p\omega_0+(1-p)\omega_1$.

The GPT formalism allows us to make probabilistic predictions of the outcomes of measurements performed on a certain state. In this context, a \textbf{measurement} is a finite collection $(e_i)_{i\in I}$ of functionals in the order interval $[0,u]$ (generically called \textbf{effects}) that add up to the order unit, i.e.\ such that $\sum_i e_i=u$ (\textbf{normalisation}). The probability of obtaining the outcome $i$ upon measuring the state $\omega\in \Omega$ is evaluated as $e_i(\omega)$. Throughout this paper, we will always make the so-called \emph{no-restriction hypothesis}, which guarantees that all normalised collections of effects identify a physically legitimate measurement~\cite{no-restriction}. We denote by $\textbf{M}$ the set of all measurements associated with a certain GPT, adding a subscript to identify it if needed. 

Equipping a vector space $V$ with a GPT structure entails selecting a special positive functional on it, i.e.\ the unit effect $u$. In turn, this special functional can be used to define a norm on the dual space $V^*$. By definition, the unit ball of this norm is the interval $[-u,u]$, and for $x^*\in V^*$ one has
\begin{equation}
\|x^*\| \coloneqq \min\left\{ t\geq 0:\ x^*\in t[-u,u] \right\} . \label{order unit norm}
\end{equation}
This choice makes $V^*$ a so-called \textbf{order unit space}~\cite{Ellis-dual-base, Ellis-66}. The corresponding Banach space structure induced on $V$ is that of a \textbf{base norm space}~\cite{Edwards-base-norm}. The norm on $V$ is given by any of the two expressions
\begin{align}
\|x\| &= \max_{x^*\in [-u,u]} |x^*(x)| \label{base norm dual}\\
&= \min\left\{ u(x_+ + x_-):\ x=x_+-x_-,\ x_\pm\geq 0 \right\} . \label{base norm}
\end{align}
The equivalence is an easy consequence of the strong duality of conic optimisation programs~\cite{BOYD}; alternatively, it can be established by checking that the convex body $K \coloneqq \conv(\Omega \cup -\Omega) \subset V$ is the unit ball for the base norm, while $\|x^*\| = \sup 
\{ |x^*(x)| \, : \, x \in K \}$ for any $x^* \in V^*$.

This more or less exhausts the description of single systems within the GPT framework. Note that dynamics is not part of this very basic picture, which is limited to so-called `prepare-and-measure procedures'. Time evolution can be accounted for within this formalism, but this goes beyond the scope of the present paper. What we will need here is instead the extension of the formalism to the case of composite systems. We will be mainly concerned with the simplest case of a bipartite system formed by two subsystems $A$ and $B$ described by local GPTs $(V_A, C_A, u_A)$ and $(V_B, C_B, u_B)$. The theory modelling the composite $AB$ will be denoted by $AB=(V_{AB}, C_{AB}, u_{AB})$. Under very reasonable assumptions~\cite{tensor-rule-1, tensor-rule-2}, the main one being that bipartite states are always uniquely determined by the statistics produced by local measurements (a principle that goes under the name of \textbf{local tomography}), one can identify $V_{AB}$ with the tensor product of the local vector spaces, i.e.\ $V_{AB}\simeq V_A\otimes V_B$. When this identification is made, one obtains also $u_{AB}=u_A \otimes u_B$.
To fully specify the joint system, one still needs to identify the cone of unnormalised states $C_{AB}$. It turns out that such a choice cannot be made a priori on the ground of some indisputable axiom, but has to be based on some information regarding the actual physics of the system. However, the operational interpretation of the theory puts some nontrivial constraints on $C_{AB}$, in the form of a lower and upper bound with respect to the inclusion relation. Namely, we have
\begin{equation}
C_A \tmin C_B \subseteq C_{AB}\subseteq C_A \tmax C_B\, ,
\label{CAB bound}
\end{equation}
where
\begin{align}
C_A \tmin C_B &\coloneqq \conv \left( C_A \otimes C_B \right) , \label{C min} \\
C_A \tmax C_B &\coloneqq \left(C_A^*\tmin C_B^*\right)^* . \label{C max}
\end{align}
The two constructions~\eqref{C min} and~\eqref{C max} are called \textbf{minimal} and \textbf{maximal tensor product}, respectively. In~\eqref{CAB bound}, the lower bound comes from the fact that any tensor product of local states must represent a valid state, while -- dually -- the fact that any tensor product of local effects must be an effect of the joint system leads to the upper bound. In~\eqref{C min} we used the notation $C_A\otimes C_B \coloneqq \left\{ x\otimes y:\, x\in C_A,\, y\in C_B \right\}$. 
%It is not difficult to check that the maximal tensor product of~\eqref{C max} can be equivalently characterised as
%\begin{equation}
%C_A \tmax C_B = \left\{ z\in V_A\otimes V_B:\ (x^* \otimes y^*)(z)\geq 0\ \forall\ x^*\in C_A^*,\ \forall\ y^*\in C_B^* \right\} .
%\end{equation}
In what follows, we will call \textbf{admissible} any composite $AB$ whose associated cone $C_{AB}$ satisfies~\eqref{CAB bound}. Also, we will denote by $A\tmin B$ and $A \tmax B$ those corresponding to the choices~\eqref{C min} and~\eqref{C max} for $C_{AB}$.

We conclude this brief presentation of the GPT formalism by discussing the two physically most relevant examples, i.e.\ classical probability theory and quantum mechanics. Classical probability theory can be viewed as the GPT 
\begin{equation}
\mathrm{Cl}_d \coloneqq \left(\R^d, \R^d_+, u \right) ,
\label{classical probability theory GPT}
\end{equation}
where $\R_+^d$ is the cone of entrywise non-negative vectors, and $u(x)\coloneqq \sum_{i=1}^d x_i$ for all $x\in \R^d$. The induced base norm coincides with the $\ell_1$-norm $\|x\|_{\ell_1}\coloneqq \sum_{i=1}^d |x_i|$. Composing classical systems is easy, for when either $C_A$ or $C_B$ is simplicial (i.e.\ a linear image of $\R_+^d$) minimal and maximal tensor product coincide.

An $n$-level quantum mechanical system is modelled by the GPT
\begin{equation}
\mathrm{QM}_{n} \coloneqq \left( \Herm{n}, \mathrm{PSD}_{n}, \tr \right) ,
\label{quantum theory GPT}
\end{equation}
where $\Herm{n}$ is the space of $n\times n$ Hermitian matrices, $\text{PSD}_n$ the cone of positive semidefinite matrices, and $\tr$ represents the trace functional. The quantum mechanical base norm is the appropriate non-commutative generalisation of the $\ell_1$-norm, i.e.\ the trace norm $\|X\|_1\coloneqq \tr \sqrt{X^\dag X}$. The base norm space of $n\times n$ Hermitian matrices endowed with the trace norm will be denoted by $\Tn$. In contrast with the classical case, for quantum mechanics composition rules become an issue. In fact, the standard quantum mechanical composition rule dictates that if $A=\text{QM}_{n}$ and $B=\text{QM}_m$ then $AB=\text{QM}_{nm}$. The corresponding cone $C_{AB}$ is well known to make both inclusions in~\eqref{CAB bound} strict.

\subsection{Tensor norms} \label{subsec tensor norms}

%We should also introduce the largest value of the $\pi/\e$ ratio. (I used $R$ at some point and will use it below but this is subject to be changed.) 
%Ideally the quantity would be given a name and a symbol that are reminiscent of the underlying operational interpretation.

We start by recalling the basic theory of tensor products of normed spaces. In what follows $B_X\coloneqq \left\{x\in X:\ \|x\|\leq 1\right\}$ will denote the unit ball of a Banach space $X$. There are at least two canonical ways in which one can construct a norm on a generic tensor product $X\otimes Y$ of finite-dimensional real Banach spaces~\cite{DEFANT, RYAN}. The \textbf{injective norm} of a tensor $z\in X\otimes Y$ is defined by the expression
\bb
\|z\|_{X \otimes_\e Y} \coloneqq \max\left\{ (x^*\otimes y^*)(z):\ x^*\in B_{X^*},\ y^*\in B_{Y^*} \right\} ,
\label{injective}
\ee
while its \textbf{projective norm} is given by
\bb
\|z\|_{X \otimes_\pi Y} \coloneqq \min\left\{ \sumno_i \|x_i\| \|y_i\|:\ z=\sumno_i x_i\otimes y_i \right\} .
\label{projective}
\ee

%\begin{note} In what follows, we will often use the simplified notation $\|\cdot\|_\e$ and $\|\cdot\|_\pi$ when there is no ambiguity on the involved spaces. \end{note}

In Subsection~\ref{subsec GPT}, we learnt that the vector space associated with a GPT carries a natural norm, i.e.\ the base norm given by~\eqref{base norm}. Since a joint system $AB$ lives on the tensor product $V_A\otimes V_B$ of the local vector spaces, it is natural to ask whether any of the above tensor norms admits an operational interpretation in this context. Indeed, it turns out that~\cite[Proposition~22]{ultimate}
\bb
\|\cdot\|_{AB}\leq \|\cdot\|_{A\tminfoot B} = \|\cdot\|_{V_A \otimes_\pi V_B}\, .
\label{pi as base norm on AB}
\ee
for all admissible composites $AB$. The last equality tells us that the projective norm corresponds to the base norm associated with the minimal tensor product of the two theories.

One may thus be led to conjecture that an analogous identity exists between $\|\cdot\|_{V_A \otimes_\e V_B}$ and $\|\cdot\|_{A\tmaxfoot B}$, but the example of two classical probability theories reveals that this is not the case. We will find an adequate operational interpretation for the injective tensor norm in the forthcoming Subsection~\ref{subsec XOR}.

The most elementary properties of injective and projective norms is perhaps the inequality
\begin{equation}
    \|\cdot\|_{X\otimes_\e Y} \leq \|\cdot\|_{X \otimes_\pi Y}\, ,
    \label{injective less than projective}
\end{equation}
valid for all $X,Y$. Moreover, since the space $X\otimes Y$ is of finite dimension, these two norms will always be equivalent, i.e.\ there will exist an optimal constant $1\leq \rho(X,Y)<\infty$ such that
\bb
\|\cdot\|_{X\otimes_\e Y} \leq \|\cdot\|_{X \otimes_\pi Y} \leq \rho(X,Y) \|\cdot\|_{X\otimes_\e Y}\, .
\ee
It is straightforward to verify that the above constant is formally given by the following optimisation:
\bb
\rho(X,Y) \coloneqq \sup_{0\neq z\in X\otimes Y} \frac{\|z\|_{X\otimes_\pi Y}}{\|z\|_{X\otimes_\e Y}}\, .
\label{rho}
\ee
For reasons that will soon become clear, in this paper we are interested in studying the range of values of the function $\rho(X,Y)$ across all pairs of spaces of fixed dimensions.

\begin{definition} \label{def pi/epsilon ratio}
The \textbf{projective/injective ratio}, or $\pi/\e$ ratio for short, is the following universal function over pairs of integers $n,m \geq 2$:
\bb
r(n,m) \coloneqq \inf_{\substack{\dim X =n \\ \dim Y =m }} \rho(X,Y)\, ,
\label{r}
\ee
where the optimisation is understood to be over all pairs of finite-dimensional Banach spaces $X,Y$ of fixed dimensions $n,m$. A slight modification of the above function~\eqref{r} yields the \textbf{symmetric projective/injective ratio}:
\bb
r_s(n) \coloneqq \inf_{\dim X=n} \rho(X,X)\, ,
\label{rs}
\ee
where $n\geq 2$ and the infimum is taken over all Banach spaces of dimension $n$.
\end{definition}

One could equally well investigate analogous quantities where the infimum in the above optimisations is replaced by a supremum, however it turns out that these can be evaluated exactly. In fact, it has been shown that~\cite[Proposition~21]{ultimate}
\bb
R(n,m) \coloneqq \sup_{\substack{\dim X=n \\ \dim Y=m}} \rho (X,Y) = \min\{n,m\}\, .
\label{R}
\ee
In light of this, in the rest of the paper we shall be concerned with the $\pi/\e$ ratios as constructed in Definition~\ref{def pi/epsilon ratio}. By taking as ansatzes $X=\ell_1^n$ and $Y=\ell_2^m$ in~\eqref{r} (and assuming without loss of generality that $n\leq m$), one can see that
\bb
1\leq r(n,m) \leq \sqrt{\min\{n,m\}} \qquad \forall\ n,m\, .
\label{elementary ineq n,m}
\ee
For an explicit proof, see the discussion preceding~\eqref{rho ell1 ell2}. To upper bound the symmetrised ratio one can take as ansatzes two copies of $\ell_1^n$, which yields the slightly worse estimate~\cite[Example~29]{ultimate} (we compute a sharper upper bound on $\rho(\ell_1^n,\ell_1^n)$, which
is equivalent to $\sqrt{\pi/2}\sqrt{n}$ as $n$ tends to infinity, in Proposition \ref{proposition:ell1-ell1})
\bb
1\leq r_s(n) \leq \rho(\ell_1^n,\ell_1^n) \leq \sqrt{2n} \qquad \forall\ n\, .
\label{elementary ineq n=m}
\ee

Note that although $r(n,n)\leq r_s(n)$ for all $n$, it may conceivably happen that $r(n,n) < r_s(n)$. In other words, it is possible that the infimum in~\eqref{r} is not achieved on two copies of the same space even when $n = m$. However, we do not know this for a fact, even when $n=2$ (cf. \eqref{bounds r(2,2)}, \eqref{bounds r_s(2)}). 

The above inequalities exhaust the elementary properties of the $\pi/\e$ ratios, and leave open many interesting questions, whose thorough investigation constitutes our main contribution. For a summary of the results we obtain on these quantities, we refer the reader to Section~\ref{sec main results}.

\subsection{XOR games} \label{subsec XOR}
 
A simple but extremely useful setting where different physical models can be studied and compared from the point of view of information processing is that defined by XOR games. In these games, a referee interacts with two players Alice and Bob, who can cooperate with each other in order to maximise their winning probability. In the classical setting, the referee chooses a pair of questions according to a publicly known distribution and sends one question to each player. Then, the players are requested to provide a one-bit answer each, and the winning condition of the game, for a given pair of questions, only depends on the parity of the answers. In the basic local setting, the players can agree in advance on a strategy for their answers but they are not allowed to communicate with each other once the game has started.

These games are arguably central in theoretical computer science, mainly because of their simplicity and broad applicability to different topics such as interactive proof systems, hardness of approximation, and the PCP theorem. In addition, XOR games have played a major role in quantum information theory since they were first considered in~\cite{CHTW04}. In fact, these games had already been implicitly considered in the context of the study of quantum nonlocality~\cite{Bell, CHSH}, by means of their equivalent formulation in terms of correlation Bell inequalities. Their systematic study was initiated by Tsirelson~\cite{Tsirelson87}. Far from being purely theoretical objects, in the last years these games have been crucial in the development of device-independent quantum cryptography and random numbers generators.

Motivated by their relevance for theory and applications, and drawing from previous works that put forth suitable quantum generalisations~\cite{Buscemi2012, RV13}, in this paper we introduce XOR games in the context of GPTs. In this more general setting, the two players' system will be described by some bipartite GPT $AB=(V_A\otimes V_B, C_{AB}, u_A\otimes u_B)$. The referee samples the questions from a finite alphabet $I$, the probability of drawing $i$ being denoted by $p_i$. The answers are represented by a collection of bits $(c_{i})_{i\in I}\in\{0,1\}^{|I|}$, while the questions are described by states $\omega_i\in C_{AB}\subset V_A\otimes V_B$. Upon being asked the question $\omega_i$, the players output answers $a\in\{0,1\}$ and $b\in\{0,1\}$, respectively, and the winning condition takes the form $a\oplus b=c_i$. The players' behaviour can be modelled by a suitable measurement $M=(g_{ab})_{ab \in \{00,01,10,11\} }\in \textbf{M}_{AB}$ over $AB$, with $g_{ab}(\omega_i)$ representing the probability that the answers $a,b$ are given when the question $\omega_i$ has been asked. We can then formalise the notion of XOR game as follows.

\begin{definition}
An \textbf{XOR game} $G$ is a quadruple $(AB, \omega, p, c)$, where: (i) $AB$ is a bipartite GPT; (ii) $\omega = (\omega_i)_{i\in I}$ is a discrete collection of states over $AB$; (iii) $p$ is a probability distribution over the set $I$; and (iv) $c=(c_i)_{i\in I}$ is a set of bits. A \textbf{strategy} for the players is a measurement $M=(g_{ab})_{ab \in \{00,01,10,11\}}$ over $AB$.
\end{definition}

The prototypical (and simplest) example of a strategy, called a \textbf{local strategy}, consists in the players performing a product measurement $(e_a \otimes f_b)_{a,b\in\{0,1\}}\in \textbf{M}_A\otimes \textbf{M}_B$. The opposite case is naturally that of a \textbf{global strategy}, corresponding to the case of Alice and Bob having access to global measurements $(g_{ab})_{a,b\in\{0,1\}}\in\mathbf{M}_{AB}$, but one can equally well consider some intermediate scenarios where the players are allowed a bounded amount of communication before they are required to output the answers. Some of these variations on the theme are examined in Appendix~\ref{app XOR games in GPTs}. From the above picture it follows that XOR games can be equivalently formulated as instances of state discrimination problems, possibly subjected to some special constraints dictated by locality.
%We will make use of this connection in a moment.

Before we delve into the study of XOR games over GPTs, let us point out some caveats in the terminology. According to the standard conventions, a \textbf{classical XOR game} is more than an XOR game played over the composite $AB=\mathrm{Cl}_{nm}$ formed by two classical GPTs $A=\mathrm{Cl}_{n}$ and $B=\mathrm{Cl}_{m}$ defined by~\eqref{classical probability theory GPT}. In fact, it is usually understood that in this case the questions are taken from the standard basis of $V_{AB}=\R^{nm}$, i.e.\ $\omega_{xy}\coloneqq v_x\otimes v_y$, where $1\leq x\leq n$ and $1\leq y\leq m$ are integers, and $v_k$ is the $k$-th vector in the standard basis of $\R^k$. In view of the perfect local distinguishability of the questions, one usually refers directly to the labels $x$ and $y$ as the questions. In compliance with the established conventions, from now on we will stick to the above definition of a classical XOR game.

We now flesh out the theory of XOR games in the GPT setting. One is usually interested in maximising the success probability of the players given a certain set of measurements they have access to. Since an XOR game can always be won with probability $1/2$ by just answering randomly, it is standard to quantify the effectiveness of the players' strategy by introducing the \textbf{bias} $\beta(G)$ of the game $G$:
\bb
\beta(G) \coloneqq P_{\text{winning}}(G) - P_{\text{losing}}(G) = 2 P_{\text{winning}}(G) - 1 \, .
\ee
It is understood that the bias depends also on the strategy adopted by the players, which we specify with a subscript. The following result yields explicit expressions for the bias corresponding to local and global strategies.

\begin{theorem} \label{theorem: bias xor games}
Consider an XOR game $G=(AB,\omega, p,c)$, where $AB=(V_A\otimes V_B, C_{AB}, u_A\otimes u_B)$. Define $z_G\coloneqq \sum_{i\in I} p_i (-1)^{c_i}\omega_i\in V_A\otimes V_B$. Then the biases corresponding to local and global strategies evaluate to
\begin{align}
\beta_{\mathrm{LO}}(G) &= \left\|z_G\right\|_{V_A\otimes_\e V_B}\, , \label{local bias} \\
\beta_{\mathrm{ALL}}(G) &= \left\|z_G \right\|_{AB}\, , \label{global bias}
\end{align}
respectively. Here, $\|\cdot\|_{AB}$ is the base norm associated with the GPT $AB$. In particular, we obtain $\beta_{\text{\emph{ALL}}}(G) \leq \left\| z_G \right\|_{V_A \otimes_\pi V_B}$ for all composites $AB$, and the bound is achieved when $AB=A\tminit B$.
\end{theorem}

\begin{proof}
It is not difficult to realise that when the players adopt global strategies, their task is equivalent to a state discrimination problem with all measurements on the composite system being available. Hence,~\eqref{global bias} follows from~\cite[Lemma~7]{ultimate}. As for~\eqref{local bias}, note that a pair of local measurements $(e_a \otimes f_b)_{a,b\in\{0,1\}}$ yields
\begin{align*}
P_{\text{winning}}(G) - P_{\text{losing}}(G) &= \sum_i p_i \left( \sum_{a,b:\, a\oplus b=c_i} (e_a\otimes f_b)(\omega_i) - \sum_{a,b:\, a\oplus b\neq c_i} (e_a\otimes f_b)(\omega_i) \right) \\
&= \sum_i p_i \sum_{a,b} (-1)^{a+b+c_i} (e_a\otimes f_b)(\omega_i) \\
&= \left( \left( \sumno_a (-1)^a e_a \right)\otimes \left( \sumno_b (-1)^b f_b\right)\right) (z_G)\, .
\end{align*}
That the optimisation over all local measurements yields~\eqref{local bias} is a consequence of the elementary fact that
\bb
\left\{ e_0-e_1:\ e_0\in [0,u_A],\ e_0+e_1=u_A \right\} = \left[-u_A, u_A\right]\, ,
\label{measurement +- unit ball}
\ee
and analogously for system $B$. The last claim follows from~\eqref{pi as base norm on AB}.
%In turn,~\eqref{measurement +- unit ball} can be proved as follows. On the one hand, since for a measurement $(e_i)_i\in\textbf{M}_A$ one has $e_i\geq 0$ for all $i$, we obtain $-e_i = -|\sigma_i| e_i\leq \sigma_i e_i \leq |\sigma_i| e_i = e_i$ and upon summation $-u_A \leq \sum_i \sigma_i e_i \leq u_A$. On the other hand, sums of this kind cover the whole unit ball of the dual norm. In fact, for $x^*\in [-u_A, u_A]$ we can define the measurement $e=\left(\frac{u_A+x^*}{2},\ \frac{u_A-x^*}{2}\right)$ and the sign vector $\sigma=(+1, -1)$. We then have $\sum_i \sigma_i e_i = \frac{u_A+x^*}{2} - \frac{u_A - x^*}{2} = x^*$, as claimed.
\end{proof}

\begin{remark}
The value of the bias under local strategies depends only on the local structure of the GPTs $A$ and $B$, and is thus independent of the particular rule we chose to compose them.
\end{remark}

Theorem~\ref{theorem: bias xor games} generalises the mathematical description of the non-entangled bias of classical and quantum XOR games~\cite{PV16, RV13}, and yields the operational interpretation of the injective norm we were seeking. In Appendix~\ref{app XOR games in GPTs} we show that this interpretation is `robust', in the sense that even allowing the players a bounded amount of communication before they output the answers does not increase the bias by more than a factor equal to the product of the dimensions of the GPTs used to carry the messages. The same type of argument also shows that not much can be gained if the players have access to a pre-shared physical system of bounded dimension. 
%In what follows, we will refer to these as \textbf{assisted strategies}.

For classical XOR games more is true: namely, Tsirelson's theorem~\cite{Tsirelson87} states that even assistance by entangled quantum states of \emph{arbitrarily large} dimension does not allow for a significant improvement over product strategies. However, this has to be regarded as a peculiar feature of quantum systems, deeply linked with the underlying Hilbert space structure through Grothendieck's inequality. It is therefore not surprising that it does not carry over to the general GPT setting. In fact, it turns out that a classical XOR game can always be won with probability $1$ if one allows for assistance from a well-chosen set of non-signalling correlations\footnote{In fact, one can pick the assisting distribution to reproduce directly the answers $a,b$ the players have to give. For a classical XOR game defined by questions $x,y$ and correct answers $c_{xy}\in \{0,1\}$, it suffices to define the assisting probability distribution by $p(ab|xy)=1/2$ if $a\oplus b=c_{xy}$ and $0$ otherwise.} (remember that non-signalling correlations can be viewed as GPTs~\cite[Section~2.3.4]{lamiatesi}).

We now illustrate the main motivation behind our investigation. 
In the scenario we consider, the referee's goal is to minimise the bias (equivalently, the winning probability) corresponding to local strategies for a fixed value of that corresponding to global strategies, effectively \emph{testing} Alice and Bob's ability to perform joint measurements on their systems. On the other hand, the two players aim to win the game with high probability while employing only local strategies. They are free to choose the local systems $A$ and $B$ of fixed local dimensions that better suit their needs, as this choice may be dictated -- among other things -- by their experimental capabilities. However, they have no control over the composite $AB$ the referee will be able to access nor over the questions they will be asked. Given these constraints, \emph{how large is the global/local bias ratio the referee can hope to achieve?}

Having specified the setting, we now proceed to perform the quantitative analysis that will ultimately lead us to identify our main object of study.
Start by observing that~\eqref{base norm} implies that every element $z\in V_A\otimes V_B$ such that $\|z\|_{AB}\leq 1$ for some legitimate composite $AB$ is such that $z=z_G$ for some game $G$ on $AB$ (with just two questions). Hence, the maximal increase in the bias of an XOR game on $AB$ when passing from the local to the global strategies is given by
\bb
\sup_{G} \frac{\beta_{\text{ALL}}(G)}{\beta_{\text{LO}}(G)} = \sup_{0\neq z\in V_A\otimes V_B} \frac{\|z\|_{AB}}{\|z\|_{V_A\otimes_\e V_B}}\, ,
\ee
where the supremum on the l.h.s.\ is over all games on a fixed composite $AB$. Since the referee is also free to choose the optimal composite $AB$ for given local systems $A$ and $B$, we can also optimise over all composition rules. This is easily done by means of~\eqref{pi as base norm on AB}, and yields
\bb
\sup_{AB, G} \frac{\beta_{\text{ALL}}(G)}{\beta_{\text{LO}}(G)} = \sup_{0\neq z\in V_A\otimes V_B} \frac{\|z\|_{V_A \otimes_\pi V_B}}{\|z\|_{V_A\otimes_\e V_B}} = \rho\left(V_A, V_B\right)\, ,
\label{ratio bias XOR = rho}
\ee
where the last step is an application of the definition~\eqref{rho}. The above equation~\eqref{ratio bias XOR = rho} is important because it connects the theory of XOR games over GPTs to that of tensor norms, and it can be used as a starting point to investigate some intrinsic aspects of the behaviour of information processing in bipartite systems.
For instance, in~\cite{ultimate} the optimal performance of data hiding against `locally constrained sets of measurements' is connected to the quantity $\sup_{A,B} \rho(V_A, V_B)$, the supremum running over all GPTs $A$ and $B$ of fixed dimensions\footnote{For convenience, here we are extending the definition of `locally constrained sets of measurements' with respect to that given in~\cite{ultimate}, including also the scenario corresponding to an XOR game.}.
%was interpreted as the ultimate bound on the efficiency of data hiding against locally constrained sets of measurements that is achievable in bipartite GPTs of fixed local dimensions. Although the supremum in~\eqref{ultimate data hiding ratio} is a priori over all pairs of \emph{base norm} spaces, it was shown in~\cite{ultimate} that one does not lose much by extending it to all pairs of Banach spaces of fixed dimensions, yielding the tight estimate~\eqref{R} on~\eqref{ultimate data hiding ratio}. This latter observation is fully general, as clarified by the results of Appendix~\ref{app base norm equivalence}.
In the setting we study here, instead, Alice and Bob's goal is to minimise the global/local bias ratio in~\eqref{ratio bias XOR = rho} by making a clever choice of $A$ and $B$ in the first place. Hence, the relevant quantity is
\bb
r_{\mathrm{bn}}(n, m) \coloneqq \inf_{\substack{\dim A = n \\ \dim B = m}} \rho\left(V_A, V_B\right) ,
\label{r bn}
\ee
the infimum running over all GPTs $A,B$ (equivalently, over all base norm spaces $V_A, V_B$) of fixed dimensions. If we find that $r_{\mathrm{bn}}(n,m)>1$ for all $n,m$, then there is a point in claiming that \emph{global strategies for XOR games perform better than local ones, independently of the underlying physical theories.} If we manage to determine the asymptotic scaling of the quantity~\eqref{r bn}, we will even be able to make this statement quantitative. Comparing~\eqref{r bn} with~\eqref{r}, it is elementary to observe that
\begin{equation}
    r_{\mathrm{bn}}(n,m)\geq r(n,m)
    \label{r bn larger than r}
\end{equation}
for all $n,m \geq 2$, as the infimum that defines the r.h.s.\ is over all pairs of Banach spaces, while that on the l.h.s.\ includes only base norm spaces. In spite of this, thanks to the result of Appendix~\ref{app base norm equivalence} we know that the two sides of~\eqref{r bn larger than r} are in fact comparable. Namely, Lemma~\ref{r vs r bn lemma} tells us that for all $n,m \geq 2$ one has
\begin{align}
r_{\mathrm{bn}}(n,m) &\leq 4\, r(n,m)\, , \label{r vs r bn same dim} \\
r_{\mathrm{bn}}(n,m) &\leq 2 + r(n-1, m-1)\, . \label{r vs r bn dim-1}
\end{align}
In light of the above equivalences, in the rest of the paper we shall study the function $r$ instead of $r_{\mathrm{bn}}$. This simplifies our investigation considerably.

Before we present our results, let us comment on the optimisation over the composition rules performed in \eqref{ratio bias XOR = rho}. While we argued above that this may not be the most natural choice, it is conceivable to consider a modified scenario where Alice and Bob fix not only the systems $A,B$, but also the composite $AB$. Instead of \eqref{ratio bias XOR = rho}, one should rather compute $\inf_{AB} \sup_G \beta_{\mathrm{ALL}}(G)/\beta_{\mathrm{LO}}(G)$, and then take also the infimum over $A$ and $B$ of fixed dimensions. Remarkably, the choice of the composite is irrelevant when either $A$ or $B$ are classical theories. In this case, we are also able to give sharp estimates of the r.h.s.\ of \eqref{ratio bias XOR = rho} (see Remarks after Proposition~\ref{proposition:ell1} and Lemma~\ref{lemma:lowerbound-ellp}).

\section{Main results} \label{sec main results}

%At this point we should have everything in hand to state the main result and see why it is of interest for both mathematicians (very general properties of tensor norms) and physicists (operational interpretation). If we send the paper to CMP or similar it may be good to state our results as soon as possible.
Throughout this section, we present our main results on the universal functions $r(n,m)$ and $r_s(n)$ introduced in~\eqref{r} and~\eqref{rs}, respectively. The discussion of the elementary properties of these objects, as conducted in Section~\ref{subsec tensor norms}, left many natural questions open. For instance, \emph{is it the case that $r(n,m)\geq c>1$ for all $n,m\geq 2$ and for some universal constant $c$?} In other words, is it true that for any pair of Banach spaces $X,Y$ of dimension at least $2$ there is a tensor $z\in X\otimes Y$ for which $\|z\|_{X \otimes_\pi Y} \geq c\|z\|_{X \otimes_\e Y}$? Looking only at large enough dimensions, \emph{how do $r(n,m)$ and $r_s(n)$ behave asymptotically in $n,m$?} 
Some insight into this latter question was provided by Pisier~\cite{Pisier80}, who showed that $r(n,m)\rightarrow \infty$ when $\min\{n,m\}\rightarrow \infty$, but with no asymptotic growth explicitly stated.

Our first result answers the first of the above questions in the affirmative.

\begin{theorem} \label{theorem: 19/18}
For any pair of Banach spaces $X$, $Y$ with $\dim X \geq 2$ and $\dim Y \geq 2$, we have $\rho(X,Y) \geq 19/18$. Equivalently, there exists a nonzero tensor $z \in X \otimes Y$ such that 
\[ \|z\|_{X \otimes_\pi Y} \geq \frac{19}{18} \|z\|_{X \otimes_\e Y}. \] Consequently, the function $r(n,m)$ defined in~\eqref{r} satisfies
\bb
r(n,m) \geq \frac{19}{18}\qquad \forall\ n,m\geq 2\, .
\ee
\end{theorem}

Theorem~\ref{theorem: 19/18} also applies to infinite-dimensional spaces and shows that the injective and projective tensor products cannot be isometric. %The fact that they can be isomorphic is highly non-trivial and follows from a construction by Pisier~\cite{Pisier}. [already mentioned before the Theorem]
Our proof of Theorem~\ref{theorem: 19/18} requires a variation on Auerbach's lemma that is susceptible to an intuitive geometrical interpretation (Lemma~\ref{lemma:hexagon-dual}). 

The following open problem asks whether the injective and projective tensor norms are always $\sqrt{2}$ apart. The value of $\sqrt{2}$ would be optimal, since $\rho(\ell_1^2,\ell_2^2) = \sqrt{2}$.

\begin{problem} \label{problem:sqrt2}
Is it true that for any pair of Banach spaces $X$, $Y$ with $\dim X, \dim Y \geq 2$, we have $\rho(X,Y) \geq \sqrt{2}$? In other words, does there always exist a nonzero tensor $z \in X \otimes Y$ such that 
\bb
\|z\|_{X \otimes_\pi Y} \geq \sqrt{2}\, \|z\|_{X \otimes_\e Y}\ ?
\ee
\end{problem}

We now move on to the analysis of the asymptotic behaviour of the function $r_s(n)$. Developing functional analysis techniques from~\cite{BM86}, we arrive at a lower estimate of $\rho(X,X)$ involving the Banach--Mazur distance of $X$ from the Hilbert space of the same dimension (Theorem~\ref{theorem: lower bound rho(X,X)}). Since by John's theorem this cannot exceed the square root of the dimension, we deduce the following estimate, which is sharp up to logarithmic factors (compare with~\eqref{elementary ineq n=m}).

\begin{theorem} \label{theorem: lower bound rs(n)}
The function $r_s$ defined in~\eqref{rs} satisfies
\bb
\sqrt{2n} \geq r_s(n) \geq c\, \frac{\sqrt{n}}{(\log n)^3}
\ee
for some universal constant $c>0$.
\end{theorem}

%Since by~\eqref{elementary ineq n=m} we know that $r_s(n) \leq \sqrt{2n}$, we obtain in particular that $r_s(n)$ is of the order of $\sqrt{n}$ for large $n$, up to logarithmic factors.

\begin{note}
Throughout the paper we will denote universal (always strictly positive) constants by $c$, $C$, $C'$ etc.  Unless explicitly indicated, these symbols do not necessarily refer to the same numerical values when they appear in different formulae.
\end{note}

The investigation of the function $r(n,m)$ poses more substantial technical hurdles. For simplicity, we start by looking at $r(n,n)$. Our main result is the following.

\begin{theorem} \label{theorem: lower bound r(n,n)}
The function $r$ defined in~\eqref{r} satisfies
\bb
\sqrt{n} \geq r(n,n) \geq c\, \frac{n^{1/6}}{(\log n)^{4/3}}
\ee
for some universal constant $c>0$. 
\end{theorem}

The above estimate shows in particular that $r(n,n)$ grows at least as a power of $n$. In turn, this implies that \emph{global strategies for XOR games are intrinsically asymptotically (much) more efficient than local ones}, which was one of our main claims. Our proof of Theorem~\ref{theorem: lower bound r(n,n)} rests upon two main ingredients: (1) a lower estimate for $\rho\left(\ell_1^n, X\right)$, $\rho\left(\ell_\infty^n, X\right)$ and $\rho\left(\ell_2^n, X\right)$ when $\dim X\geq n$ (Lemma~\ref{lemma:lowerbound-ellp}), which can be handled using known facts about $p$-summing norms; and (2) a `trichotomy theorem' inspired by previous results in~\cite{Rudelson95, STJ09}, which states that every Banach space hosts sufficiently well-behaved subspaces on which the norm is similar enough to either an $\ell_1$-norm, or an $\ell_\infty$-norm, or a Euclidean norm (Theorem~\ref{theorem: trichotomy}).

Due to the technical complexity of managing many different estimates simultaneously, for the case $n\neq m$ we could not obtain an exponent as good as $1/6$. However, we were nevertheless able to ensure that there is power law scaling in $\min\{n,m\}$.

\begin{theorem} \label{theorem: lower bound r(n,m)}
For all $n,m\geq 2$, the function $r(n,m)$ satisfies
\bb
\sqrt{\min\{n,m\}} \geq r(n,m) \geq c\, \frac{\, \min\{n,m\}^{1/8}\, }{\log \min\{n,m\}}
\ee
for some universal constant $c>0$.
\end{theorem}

The exponents we obtained in Theorems~\ref{theorem: lower bound r(n,n)} and~\ref{theorem: lower bound r(n,m)} are unlikely to optimal. We present the following problem concerning the scaling of the projective/injective ratio.

\begin{problem} \label{problem:asymptotic}
Does there exist a universal constant $c>0$ such that
\bb
r(n,m)\geq c \min\{n,m\}^{1/2}
\label{eq: r scaling conjecture 1}
\ee
for all positive integers $n,m$? In other words, for all pairs of finite-dimensional Banach spaces $X$, $Y$, does there exist a nonzero tensor $z\in X\otimes Y$ such that
\bb
\|z\|_{X\otimes_\pi Y} \geq c \min\{\dim X, \dim Y\}^{1/2}\, \|z\|_{X\otimes_\e Y}\ ?
\label{eq: r scaling conjecture 2}
\ee
\end{problem}

While the value of $r(n,m)$ grows with the dimension, new phenomena appear when considering infinite-dimensional spaces. Indeed, a famous construction by Pisier~\cite{Pisier}, solving negatively a conjecture by Grothendieck, entails that there exists an infinite-dimensional Banach space $X$ such that $\rho(X,X)<\infty$. Using the information from Lemma~\ref{2-isomorphism base norm space lemma}, we conclude that the same behaviour occurs in the realm of GPTs: there exist infinite-dimensional GPTs $A$, $B$ such that local and global strategies are equivalent up to a universal constant in any possible composite $AB$.

The study of asymptotic behaviours in the general setting should not distract us from the fact that certain GPT models are of prime importance because of their compliance with known physics. Therefore, one of our results is the determination of the quantity $\rho(X,Y)$ when $X,Y$ are the base norm spaces corresponding to quantum mechanical systems, i.e.\ the Banach spaces $\Tk$. The following constitutes a notable improvement over~\cite[Lemma~20]{Brandao-area-law},~\cite[Theorem~15]{VV-dh} and~\cite[Eq.~(72)]{ultimate}, as detailed below.

\begin{theorem}\label{theorem: rho for S1}
Denoting by $\Tk$ the space of $k\times k$ Hermitian matrices endowed with the trace norm, the best constant of domination of $\|\cdot\|_{X \otimes_\e Y}$ over $\|\cdot\|_{X \otimes_\pi Y}$ on $\Tn\otimes \Tm$ satisfies
\bb
c \min\{n,m\}^{3/2} \leq \rho\left( \Tn,  \Tm \right) \leq C \min\{n,m\}^{3/2}
\label{eq: rho S1 scaling}
\ee
for some constants $C$, $c>0$. More precisely, we have the following estimate for the upper bound in the above relation:
\bb
\rho\left( \Tn,  \Tm \right) \leq 4 \min\{n,m\}^{3/2} - 2\sqrt2 (\sqrt2-1)\sqrt{\min\{n,m\}}.
\label{eq: rho S1 upper precise}
\ee
\end{theorem}

\begin{note}
When comparing~\eqref{eq: rho S1 scaling} with the other estimates on $r(n,m)$ that we presented throughout this section, one should remember that the dimension of the space $\Tn$ is $n^2$ rather than $n$. 
Hence, curiously, for $X=\Tn$ the quantity $\rho(X,X)\approx n^{3/2}$ is of the same order as the geometric mean between the theoretical minimum $r_s(n^2)\approx n$ and the absolute maximum $R(n^2,n^2)=n^2$.
\end{note}

Although we will not consider complex spaces in this work, it is worth mentioning that the same estimate holds (with a slight modification in the constants) if the real space $\Tk$ is replaced by the complex space $S_1^k$ of $k\times k$ matrices endowed with the trace norm.

Among other things, the above Theorem~\ref{theorem: rho for S1} enables us to give a new upper bound on the maximal efficiency of quantum mechanical data hiding against local measurements, quantitatively encoded in the data hiding ratio function $R_{\mathrm{LO}}(n,m)$ defined in~\cite{ultimate}.

\begin{corollary} \label{cor data hiding}
For all Hermitian matrices $Z$ acting on a bipartite system $\mathbf{C}^n\otimes \mathbf{C}^m$, we have
\bb
\|Z\|_{\Tn\otimes_\e \Tm} \leq \|Z\|_{\mathrm{LO}} \leq \|Z\|_1 \leq \|Z\|_{\Tn\otimes_\pi \Tm} \leq 4\min\{n,m\}^{3/2} \|Z\|_{\Tn\otimes_\e \Tm}\, , 
\label{eq: LO norm QM}
\ee
where $\|\cdot\|_{\mathrm{LO}}$ is the distinguishability norm under local measurements~\cite{VV-dh}. In other words, the data hiding ratio against local measurements can be upper bounded as
\bb
R_{\mathrm{LO}}(n,m)\leq 4 \min\{n,m\}^{3/2}\, .
\label{data hiding ratio LO QM}
\ee
\end{corollary}

The above result improves upon several previously known estimates. In~\cite[Lemma~20]{Brandao-area-law}, an inequality analogous to~\eqref{data hiding ratio LO QM} was shown, that featured an exponent $2$ on the r.h.s.; moreover, the relation proven in~\cite[Theorem~15]{VV-dh} implies that $R_{\mathrm{LO}}(n,m)\leq \sqrt{153\, n m}$, which can be worse than~\eqref{data hiding ratio LO QM} e.g.\ when $n^2 \leq m$ or $m^2\leq n$.

\section{First bounds on the \texorpdfstring{$\pi/\e$}{pi/epsilon} ratio}

\label{sec first bounds}

\subsection{Some notions of functional analysis}

We start by reminding the reader of some facts in elementary linear algebra. Given a pair of finite-dimensional vector spaces $X$, $Y$, there is a canonical isomorphism between the tensor product space $X\otimes Y$ and the space of linear maps $X^*\rightarrow Y$. We will write this correspondence as
\begin{equation}
    X\otimes Y \ni z\longmapsto \op{z} \in \pazocal{L}(X^*,Y)\, .
    \label{tensor-operator}
\end{equation}
Note that one has $\op{F(z)}= \op{z}\,^*$, where on the l.h.s.\ we have the flip operator $F:X\otimes Y\rightarrow Y\otimes X$ defined by $F(x\otimes y)\coloneqq y\otimes x$, while on the r.h.s.\ $(\cdot)^*$ stands for the adjoint (transposition) operation $\pazocal{L}(X^*,Y)\rightarrow\pazocal{L}(Y^*, X)$. It is also easy to see that one has
\begin{equation}
w(z) = \tr \left[\op{z}\,^* \op{w}\right] = \tr \left[ \op{z}\, \op{w}\,^*\right] \qquad \forall\ z\in X\otimes Y,\quad \forall\ w\in (X\otimes Y)^*=X^*\otimes Y^*\, ,
\label{evaluation = trace}
\end{equation}
where $\tr$ denotes the trace.

In this paper we are interested in tensor products of finite-dimensional real Banach spaces, so from now on $X$ and $Y$ will denote a pair of such objects. We already encountered the concepts of injective and projective tensor products (see~\eqref{injective} and~\eqref{projective}). Below we discuss some elementary properties of these constructions.
For a start, injective and projective norm are dual to each other, in the sense that
\bb
\|\cdot\|_{(X\otimes_\e Y)^*} = \|\cdot\|_{X^*\otimes_\pi Y^*}\, .
\label{duality epsilon pi}
\ee
By means of the correspondence~\eqref{tensor-operator}, it is possible to translate tensor norms into the language of operators. One has
\begin{align}
\|z\|_{X\otimes_\e Y} &= \left\|\op{z}:X^*\rightarrow Y\right\| , \label{injective = operator norm}\\
\|z\|_{X\otimes_\pi Y} &= \left\|\op{z}:X^*\rightarrow Y\right\|_N , \label{projective = nuclear norm}
\end{align}
where $\|\cdot\|_N$ is the so-called \textbf{nuclear norm}~\cite[Section~8]{TOMCZAK}.

The \textbf{Banach--Mazur distance} between two normed spaces $X$, $Y$ with the same finite dimension is defined
as~\cite{TOMCZAK}
\begin{equation}
\dbm (X,Y) \coloneqq \inf_u \left\{ \| u : X \to Y \| \cdot \| u^{-1} : Y \to X \| \right\}, \label{Banach-Mazur distance}
\end{equation}
where the infimum is taken over invertible linear maps $u$ from $X$ to $Y$. It satisfies the multiplicative version of the triangle inequality, i.e.\  $\dbm (X,Z) \leq \dbm (X,Y)\dbm (Y,Z)$. Another elementary property is the fact that $\dbm (X,Y)=\dbm (X^*,Y^*)$. 
When $\dbm (X,Y) \leq \lambda$, we say that $X$ is {\bfseries $\lambda$-isomorphic} to $Y$. Similarly, $X$ is {\bfseries $\lambda$-Euclidean} if it is $\lambda$-isomorphic to $\ell_2^{\dim X}$.

The Banach--Mazur distance only makes sense for a pair of spaces of equal dimension. When $X$, $Y$ are normed spaces such that $\dim X \leq \dim Y$, one may define as a substitute the \textbf{factorisation constant} of $X$ through $Y$ as
\begin{equation}
\f (X,Y) \coloneqq \inf_{u,v} \left\{ \| u : X \to Y \| \cdot \| v : Y \to X \|  \, : \, vu = \id_X \right\}. \label{factorisation constant}
\end{equation}
Clearly, $\f (X,Y)$ is finite if and only if $\dim X\leq \dim Y$ (otherwise the above infimum is over an empty set). In order to circumvent this restriction, it is sometimes relevant to consider a relaxed version of the above quantity, where we allow the factorisation to be realised only after averaging. This leads to the definition of the \textbf{weak factorisation constant} as 
\begin{equation}
\wf (X,Y) \coloneqq \inf_{u,v} \left\{ \E \left[ \| u : X \to Y \| \cdot \| v : Y \to X \|  \right] \, : \, 
\E \left[ vu \right] = \id_X \right\},
\label{weak factorisation constant}
\end{equation}
where now the infimum is taken over pairs of operator-valued random variables $(u,v)$. We get the inequality $\wf (X,Y) \leq \f (X,Y)$ by restricting the infimum to constant random variables. Note also that the quantity $\wf (X,Y)$ is well-defined without any restriction on the dimensions of $X$ and $Y$. It is easy to verify that the factorisation constants dualise, in the sense that
\begin{align}
    \f (X,Y) &= \f (X^*, Y^*)\, , \label{f dualises}\\
    \wf (X,Y) &= \wf (X^*, Y^*)\, . \label{wf dualises}
\end{align}

We may also consider a symmetric variant of the weak factorisation constant, called
the {\bfseries weak Banach--Mazur distance} and defined as
\bbb
\weakd(X,Y) = \max\{ \wf(X,Y), \wf(Y,X)\}\, .
\eee

The family of all equivalence classes of $n$-dimensional normed spaces up to isometries can be turned into a compact metric space by the introduction of the distance $\log \dbm$. We now review some classical facts about the geometry of such space, called the \textbf{Banach--Mazur compactum} of dimension $n$. For a more complete introduction, we refer the reader to the excellent monograph~\cite{TOMCZAK}. A general upper bound valid for any $n$-dimensional normed space is the estimate following from John's theorem (Theorem~\ref{theorem:john})
\begin{equation}
\dbm (X,\ell_2^n) \leq \sqrt{n}.
\label{John estimate}
\end{equation}
As a consequence, for any pair of $n$-dimensional spaces $X$, $Y$, we have
\begin{equation} \label{eq:diameterBM-upper bound} \dbm (X,Y) \leq n . \end{equation}
This bound is essentially sharp: Gluskin proved, via a random construction, the existence of spaces $X$, $Y$ such
that $\dbm (X,Y) \geq cn$ for some $c>0$~\cite{Gluskin1981}. However, the estimate~\eqref{eq:diameterBM-upper bound} can be improved
in many specific cases. In particular, a question that is of relevance to us is that of the distance between a space and its dual: it was proved by Bourgain and Milman~\cite{BM86} that whenever $\dim(X)=n$, we have
\begin{equation} \label{eq:Bourgain-Milman}\dbm (X,X^*) \leq C n^{5/6} \log^C n. \end{equation}
In a similar vein,~\eqref{eq:diameterBM-upper bound} can be improved if we switch to the weak Banach--Mazur distance:
a result by Rudelson~\cite{Rudelson95} asserts that 
\begin{equation} \label{eq:Rudelson} \weakd (X,Y) \leq C n^{13/14} \log^C n \end{equation}
whenever $\dim(X)=\dim(Y)=n$. 
We also point out that the exact diameter of the Banach--Mazur compactum is known in dimension $2$: namely,
\begin{equation}\label{Diameter in dim 2}
\max_{\substack{\dim X = 2 \\ \dim Y = 2}} \dbm (X,Y) = \frac{3}{2}\, ,
\end{equation}
and the equality is achieved iff the unit balls of $X$ and $Y$ are the images of a square and a regular hexagon through a linear (invertible) map~\cite{Stromquist81}.

\subsection{Basic properties}

We start the investigation of the elementary properties of the parameter $\rho(X,Y)$ by presenting a simple calculation in the case $X=\ell_1^n$, $Y=\ell_2^m$, where $n\leq m$ (we point that the calculation can be rephrased in terms of $1$-summing norms, see Proposition \ref{proposition:ell1-X}). For a matrix $z\in \ell_1^n\otimes \ell_2^m$, we write
\begin{align*}
\|z\|_{\ell_1^n\otimes_\e \ell_2^m} &= \left(\sup_{\sigma\in \{\pm 1\}^n} \sumno_{j=1}^m \left(\sumno_{i=1}^n z_{ij} \sigma_i \right)^2\right)^{1/2} \\
&\geq \left(\E_\sigma \sumno_{j=1}^m \left(\sumno_{i=1}^n z_{ij} \sigma_i \right)^2\right)^{1/2} \\
&= \left(\sumno_{j=1}^m \sumno_{i=1}^n z_{ij}^2\right)^{1/2} \\
&\geq \frac{1}{\sqrt{n}}\sumno_{i=1}^n \left( \sumno_{j=1}^m z_{ij}^2\right)^{1/2} \\
&= \frac{1}{\sqrt{n}} \|z\|_{\ell_1^n\otimes_\pi \ell_2^m}\, ,
\end{align*}
where to obtain the first inequality we randomised over $\sigma$, assuming that $\sigma_1, \ldots, \sigma_n$ are i.i.d.\ $\pm 1$ Bernoulli random variables. The above computation shows that $\|z\|_{\ell_1^n\otimes_\pi \ell_2^m}\leq \sqrt{n} \|z\|_{\ell_1^n\otimes_\e \ell_2^m}$, implying that $\rho(\ell_1^n, \ell_2^m)\leq \sqrt{n}$. That this upper bound is in fact tight can be seen by considering the matrix $z$ with entries $z_{ij} = \delta_{i,j}$ (i.e.\ the identity if $m=n$). We conclude that
\bb
\rho(\ell_1^n, \ell_2^m) = \sqrt{n} \qquad \forall\ n\leq m\, ,
\label{rho ell1 ell2}
\ee
entailing the upper bound \eqref{elementary ineq n,m} on $r(n,m)$.

We now move on to investigating some more general properties of the function $\rho(X,Y)$. We start by stating a very useful reformulation in terms of operators rather than tensors.

\begin{lemma} \label{lemma: rho and trace(uv)}
For any pair of finite-dimensional normed spaces $X$, $Y$, we have
\begin{equation} \label{eq:reformulation-trace} \rho(X,Y) = \sup \left\{ \tr(vu) \, : \ \| u : X \to Y^* \| \leq 1, \ \| v : Y^* \to X \| \leq 1 \right\}. \end{equation}
\end{lemma}

\begin{proof}
The statement follows from the properties of injective and projective norms under the correspondence~\eqref{tensor-operator}. We have 
\begin{align*}
\rho(X,Y) &= \sup_{0\neq z\in X\otimes Y} \frac{\|z\|_{X\otimes_\pi Y}}{\|z\|_{X\otimes_\e Y}} \\
&\texteq{1} \sup_{0\neq z\in X\otimes Y}\sup_{0\neq w\in X^*\otimes Y^*}\frac{w(z)}{\|w\|_{X^*\otimes_\e Y^*} \|z\|_{X\otimes_\e Y}} \\
&= \sup \left\{ w(z)\,:\ \|w\|_{X^*\otimes_\e Y^*}\leq 1,\ \|z\|_{X\otimes_\e Y}\leq 1 \right\} \\
&\texteq{2} \sup\left\{\tr\left[ \left(\op{z}\right)^* \op{w} \right]\, : \ \|\op{w}:X\to Y^*\|\leq 1,\ \|\op{z}: X^*\to Y\|\leq 1 \right\} \\
&\texteq{3} \sup\left\{\tr\left[ vu \right]\, : \ \|u:X\to Y^*\|\leq 1, \ \|v:Y^*\to X\|\leq 1 \right\} .
\end{align*}
The above passages can be justified as follows: 1: we used the duality relation~\eqref{duality epsilon pi}; 2: we applied~\eqref{evaluation = trace},~\eqref{injective = operator norm} and~\eqref{projective = nuclear norm}; 3: we just renamed $v\coloneqq \left(\op{z}\right)^*$ and $u\coloneqq \op{w}$.
\end{proof}

The following proposition gathers several estimates of the function $\rho(X,Y)$.

\begin{proposition} \label{proposition: rho and distance}
Let $X$, $X'$ and $Y$ be finite-dimensional normed spaces. Then
\begin{align}
 \rho(X,Y) &= \rho(Y,X) = \rho(X^*,Y^*) = \rho(Y^*,X^*)\, , \label{eq: rho dual and exchange} \\
 \rho(X,X^*) &= \dim (X)\, , \label{eq: rho XX*} \\
 \rho(X',Y) &\leq \wf (X',X) \rho(X,Y)  \label{eq: rho heredity wf} \\
 & \leq  \f(X',X) \rho(X,Y) \, , \label{eq: rho heredity f} \\
 \rho(X',Y) & \leq \dbm(X',X) \rho(X,Y)  \ \ \ (\textnormal{assuming } \dim (X) = \dim(X') ), \label{eq: rho heredity d}\\
 \rho(X,Y) &\geq \frac{\dim X}{\wf (X,Y^*)}\, , \label{eq: lower bound wf} \\
 \rho(X,Y) &\leq \min\left\{\dim (X), \dim(Y)\right\} . \label{eq: rho upper bound}
\end{align}
In particular, when $\dim(X) =\dim(Y)=n$ we have
\begin{equation}
\rho(X,Y) \geq \frac{n}{\weakd (X,Y^*)} \geq \frac{n}{\dbm (X,Y^*)}\, . \label{eq: lower bound d}
\end{equation}
\end{proposition}

\begin{proof}
The identities~\eqref{eq: rho dual and exchange} can be obtained for instance from~\eqref{eq:reformulation-trace} by exchanging the role of $u$ and $v$ and/or taking their duals $u^*, v^*$. Remember that one has $\tr (uv) = \tr (vu) = \tr (u^* v^*) = \tr (v^*u^*)$.
Also~\eqref{eq: rho XX*} is elementary, and follows by taking $u=v=\id_X$ in~\eqref{eq:reformulation-trace}.

To show~\eqref{eq: rho heredity wf}, pick some operators $u:X'\to Y^*$ and $v:Y^*\to X'$ of norm no larger than $1$, and consider random variables $u':X'\to X$ and $v':X\to X'$ such that $\E[v'u']=\id_{X'}$. Since for all realisations of $u'$ and $v'$ the operator $\frac{uv'}{\|v'\|}:X\to Y^*$ has norm no larger than $1$, and an analogous reasoning holds for $\frac{u'v}{\|u'\|}:Y^*\to X$, using~\eqref{eq:reformulation-trace} we deduce that 
\begin{equation*}
    \frac{\tr (u'vuv')}{\|u'\|\|v'\|} \leq \rho(X,Y)\, .
\end{equation*}
Then we can write
\begin{align*}
    \tr (vu) &= \tr \left( vu \E[v'u']\right) \\
    &= \E \left[ \tr \left(vuv'u'\right)\right] \\
    &= \E \left[ \tr\left( u'vuv'\right)\right] \\
    &= \E \left[\|u'\|\|v'\| \frac{\tr(u'vuv')}{\|u'\|\|v'\|}\right] \\
    &\leq \rho(X,Y) \E \left[\|u'\|\|v'\|\right] .
\end{align*}
Taking the supremum over $u,v$ and the infimum over $u',v'$ subjected to the above constraints, and using~\eqref{weak factorisation constant} and~\eqref{eq:reformulation-trace}, we finally obtain~\eqref{eq: rho heredity wf}. Since $\wf(X',X) \leq \f(X',X)$,~\eqref{eq: rho heredity f} follows as well. To prove~\eqref{eq: rho heredity d}, note that $\dbm(X,X')=\f(X,X')$ whenever $\dim(X)=\dim(X')$. Note that~\eqref{eq: lower bound wf} also follows immediately:
\begin{align}
    \dim X &\texteq{1} \rho(X,X^*) \\
    &\textleq{2} \wf (X,Y^*) \rho(Y^*,X^*) \\
    &\texteq{3} \wf (X,Y^*) \rho(X,Y)\, .
\end{align}
where: 1: follows from~\eqref{eq: rho XX*} and~\eqref{eq: rho dual and exchange}; 2: is an application of~\eqref{eq: rho heredity wf}; and 3: is again a consequence of~\eqref{eq: rho dual and exchange}.

The estimate~\eqref{eq: rho upper bound} was proved in~\cite[Proposition~21]{ultimate} (compare with~\eqref{R}) as a consequence of Auerbach's lemma~\cite[Vol.~I, Sec.~1.c.3]{LINDENSTRAUSS}. An alternative proof can be given using the following fact from linear algebra (left as an exercise for the reader): every linear map $w$ on $\R^n$ that is a contraction for some norm satisfies $\tr(w)\leq \mathrm{rank}(w)$. To recover~\eqref{eq: rho upper bound}, apply this observation to the composite map $w=vu$, with $u$, $v$ as in~\eqref{eq:reformulation-trace}.
\end{proof}

\begin{remark}
It follows in particular from~\eqref{eq: rho heredity f} that any upper bound on $\rho(X,Y)$ is also valid for $1$-complemented subspaces of $X$ and $Y$. Recall that a subspace $X' \subseteq X$ is $\boldsymbol{\lambda}$-\textbf{complemented} if there is a surjective projection $P:X \to X'$ with $\|P\| \leq \lambda$. The complementation hypothesis cannot be omitted. To give a concrete example, consider a $cn$-dimensional subspace $X \subseteq \ell_1^{n}$ with $\dbm (X,\ell_2^{cn}) \leq C$ (existence of such a subspace is well known and follows for example from Dvoretzky--Milman theorem (Theorem~\ref{theorem: dvoretzky}) since $k_*(\ell_1^n) \geq cn$): we have $\rho(X,X) = \Theta(n)$ while $\rho(\ell_1^{n},\ell_1^{n}) = \Theta(\sqrt{n}).$
\end{remark}

\begin{remark}
From~\eqref{eq: rho heredity d} we see in particular that the function $\rho(\cdot, \cdot)$ defined on the product of the Banach--Mazur compacta of dimensions $n$ and $m$ is continuous (with respect to the product metric). In particular, this implies that the infima in~\eqref{r} and~\eqref{rs} are always achieved. We will make use of this fact without further mention in what follows.
\end{remark}

We point out that weaker versions of Theorems~\ref{theorem: lower bound rs(n)} and~\ref{theorem: lower bound r(n,n)} follow easily by combining Proposition~\ref{proposition: rho and distance} and `off-the-shelf' results. More precisely, the lower bound $r_s(n) \geq cn^{1/6}/(\log n)^C$ is an immediate consequence of~\eqref{eq: lower bound d} and~\eqref{eq:Bourgain-Milman}, and the lower bound $r(n,n) \geq cn^{1/14}/(\log n)^C$ follows from~\eqref{eq: lower bound wf} and~\eqref{eq:Rudelson}. Interestingly, the special case $n=m$ of Problem~\ref{problem:asymptotic} would follow from~\eqref{eq: lower bound d} if one could prove Rudelson's conjecture that $\wf(X,Y)\leq C \sqrt{n}$ for all $n$-dimensional Banach spaces $X,Y$.

\medskip

In the case where one of the spaces is $\ell_1^n$, the quantity 
$\rho(X,Y)$ can be rephrased in terms of $1$-summing norms (the
quantities $\pi_1^{(n)}(u)$ and $\pi_1(u)$ are defined in 
Appendix \ref{app p-summing}).

\begin{proposition} \label{proposition:ell1-X}
For every finite-dimensional normed space $X$, we have
\[ \rho(\ell_1^n,X) = \pi_1^{(n)}(\id_X) \leq \pi_1(\id_X) .\]
\end{proposition}

\begin{proof}
Using Lemma \ref{lemma: rho and trace(uv)}, we have
\begin{equation} \label{eq:rho ell1 X} \rho(\ell_1^n,X) = \rho(X,\ell_1^n) = \sup \{ \tr (vu) \st \| u : X \to \ell_{\iy}^n \| \leq 1, \ \|v : \ell_{\iy}^n \to X \| \leq 1 \}. \end{equation}
We rewrite the norms which appear in \eqref{eq:rho ell1 X} in a more tangible way. If $v : \ell_{\iy}^n \to X$ and $v_i=v(e_i)$, then 
\begin{align}\|v\|&=  \sup \left\{ \left\|\sum_{i=1}^n \e_i v_i \right\| \st \e_i=\pm 1 \right\} \\
&=  \sup \left\{ \sum_{i=1}^n \big|\scalar{f}{v_i}\big| \st f\in X^*,\ \|f\|\leq 1 \right\}. \nonumber
\end{align}

Next, if $u : X \to \ell_\iy^n$ is given by  $x \to (\scalar{f_i}{x})_{i=1}^n$ for some $f_i \in X^*$, then $\|u\|= \max_i \|f_i\|$. 
Finally, if $u$ and $v$ are as above, then  $\tr(vu)  =\tr(uv)  = \sum_{i=1}^n \scalar{f_i}{v_i}$.  Combining these we are led to 
\bb \label{ell-1-N-X} \rho(\ell_1^n,X) = \sup \left\{ \sum_{i=1}^n \|v_i\| \, : \  v_i \in X,\ \max_{\e_i=\pm 1} \left\|\sum_{i=1}^n \e_i v_i\right\|  \leq 1 \right\}. 
\ee
Comparing with \eqref{def p-summing} and using the relation
\[ \max_{\e_i=\pm 1} \left\|\sum_{i=1}^n \e_i v_i\right\| = 
\sup_{\phi \in B_{X^*}} \sum_{i=1}^n |\phi(v_i)| \]
shows that $\rho(\ell_1^n,X) = \pi_1^{(n)}(\id_X)$ 
(the general inequality $\pi_1^{(n)}(\cdot) \leq \pi_1(\cdot)$ is immediate 
from the definitions).
\end{proof}

With this connection at hand, we are able to give an improved upper bound on the parameter $r_s(n)$.

\begin{proposition} \label{proposition:ell1-ell1}
For every $n \geq 1$, we have
\[ r_s(n) \leq \rho(\ell_1^n,\ell_1^n) \leq \pi_1(\id_{\ell_1^n}) = \frac{n}{ \E \left| \sum_{i=1}^n \e_i \right| } \stackrel{n\to\iy}{\sim} \sqrt{\frac{\pi n}{2}}, \]
where $\e_1,\ldots, \e_n\in \{\pm 1\}$ are independent random variables with $\P(\e_i=1)=\P(\e_i=-1)=1/2$.
\end{proposition}

\begin{proof}
This is a simple consequence of Proposition \ref{proposition:ell1-X} and \eqref{eq:1-summing-ell1}.
\end{proof}

\subsection{Universal lower bounds}

In this section we prove that the injective and projective tensor products of any two Banach spaces cannot be isometric, unless one of them is $1$-dimensional. Going further, Problem~\ref{problem:sqrt2} asks whether the injective and projective norms are always at least $\sqrt{2}$ apart. As a partial answer, we present a proof in the special case when one of the spaces is $\ell_{\iy}^n$ (Proposition~\ref{proposition:ell1}). Then, we solve a weaker version of Problem~\ref{problem:sqrt2}, with $\sqrt{2}$ replaced by the value $19/18$ (Theorem~\ref{theorem: 19/18}). Finally, we discuss the special case when both dimensions are equal to $2$.

\begin{proposition} \label{proposition:ell1}
If $Y$ is a Banach space with $\dim(Y) \geq 2$, then for any $n \geq 2$,
\[ \rho(\ell_{\iy}^n,Y) \geq \sqrt{2}. \]
\end{proposition}

\begin{proof}
Since $\ell_{\iy}^2$ is $1$-complemented in $\ell_{\iy}^n$, in view of~\eqref{eq: rho heredity f}, it suffices to consider $n=2$. In that case, there are explicit formulas for both the projective and injective norms: for any $z=e_1 \otimes y_1 + e_2 \otimes y_2 \in \ell_\infty^2 \otimes Y$, we have \[ \|z\|_{\ell_{\iy}^2 \otimes_\pi Y}= \frac{1}{2} \left( \|y_1+y_2\| + \|y_1-y_2\|\right) \ \ \textnormal{and} \ \ \|z\|_{\ell_{\iy}^2 \otimes_\e Y}= \max \left\{ \|y_1\|,\|y_2\| \right\} . \]
It remains to justify that any Banach space $Y$ contains two vectors $y_1$ and $y_2$ such that
\begin{equation} \label{eq: modulus convexity} 
 \|y_1+y_2\| + \|y_1-y_2\| \geq 2\sqrt{2} \max \left\{ \|y_1\|,\|y_2\| \right\}. \end{equation}
This follows from properties of the so-called \textbf{modulus of uniform convexity} of $Y$, a real function defined for $\e \in [0,2]$ by 
\[ \delta_Y(\e) = \inf \left\{ 1 - \frac{\|y_1+y_2\|}{2} \st \|y_1\|=\|y_2\|=1, \ \|y_1-y_2\| \geq \e \right\}.\]
It is known~\cite{Nordlander60} that for any Banach space $Y$ and any $\e \in [0,2]$, we have $\delta_Y(\e) \leq 1-\sqrt{1-{\e^2}/{4}}$ (the value obtained for a Euclidean space). 
Applying this inequality with $\e=\sqrt{2}$ show the existence of unit vectors $y_1$, $y_2$ such that $\|y_1-y_2\| = \sqrt{2}$ and $\|y_1+y_2\| \geq \sqrt{2}$, and therefore~\eqref{eq: modulus convexity} is satisfied.
\end{proof}

\begin{remark}
An immediate consequence of Proposition~\ref{proposition:ell1} is that there is always a gap at least as large as $\sqrt2$ between local and global bias for XOR games played over a system $AB$ in which either $A$ or $B$ is a classical theory \eqref{classical probability theory GPT}.
\end{remark}

\begin{figure}[htbp] \begin{center}
\begin{tikzpicture}[scale=2]
	\coordinate (a) at (0,1);
    \coordinate (b) at (2/3,2/3);
	\coordinate (c) at (1,0);
	\coordinate (d) at (0,-1);
	\coordinate (e) at (-2/3,-2/3) ;
	\coordinate (f) at (-1,0) ;
	\coordinate (A) at (1,-1);
	\coordinate (B) at (-1,-1);
	\coordinate (C) at (-1,1);
	\coordinate (D) at (1,1) ;
	\draw[fill=gray!30] (A)--(B)--(C)--(D)--(A) ;
	\draw[fill=gray!90] (a)--(b)--(c)--(d)--(e)--(f)--(a) ;
	\draw (-.8,.8) node {$S$};
	\draw (.3,.2) node {$H$};
	\end{tikzpicture} \end{center}
    \caption{Any centrally symmetric planar convex body has a linear image $K$ satisfying $H \subseteq K \subseteq S$, where $S$ is the square $[-1,1]^2$ and $H$ is the hexagon with vertices $(\pm 1,0)$, $(0,\pm 1)$ and $(\pm 2/3, \pm 2/3)$.}
\label{figure:hexagon}
\end{figure}
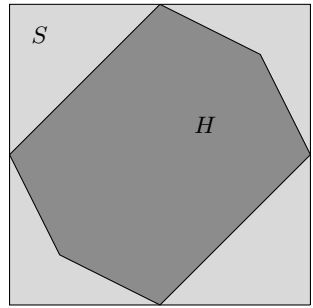

We now proceed to establish the universal lower bound $r(n,m)\geq 19/18$, formalised earlier as Theorem~\ref{theorem: 19/18}. Our main technical tool is the following variant of Auerbach's lemma, illustrated in Figure~\ref{figure:hexagon}.

\begin{lemma} \label{lemma:hexagon-dual}
Let $X$ be a Banach space of dimension at least $2$. Then there exist vectors $e_1$, $e_2 \in X$, $e_1^*$, $e_2^* \in X^*$ such that for any $i,j \in \{1,2\}$ we have $\|e_i\|_X = \|e_j^*\|_{X^*} = 1$, $e_j^*(e_i) = \delta_{i,j}$, and moreover 
\[ \|e_1 + e_2 \|_X \leq 3/2.\]
\end{lemma}

\begin{proof}
It is enough to prove the lemma when $\dim X =2$, since the general case follows by considering any $2$-dimensional subspace $Y \subseteq X$ and extending the linear forms.

Suppose now $\dim X=2$. Without loss of generality we may assume that $X = (\R^2,\|\cdot\|_X)$, and identify as well $X^*$ with $\R^2$. By applying a suitable linear transformation, we may assume that the variational problem $\max \{ |\det(f,g)| : \, f,g \in B_{X^*} \}$ is achieved when $(f,g)=(e_1,e_2)$, the canonical basis. It is clear that $\|e_1\|_{X^*} = \|e_2\|_{X^*}  =1 $, and one checks that $\|e_1\|_X=\|e_2\|_X=1$. Let us show this explicitly for $e_1$. On the one hand, $1 = \scalar{e_1}{e_1} \leq \|e_1\|_{X^*} \|e_1\|_X = \|e_1\|_X$. On the other hand, if $\|e_1\|_X>1$ one could find a vector $f'\in B_{X^*}$ such that $\scalar{f'}{e_1}>1$, which would imply that $|\det (f',e_2)| >1 = |\det (e_1,e_2)|$, in contradiction with the assumption that the pair $(e_1,e_2)$ achieves the maximum in the above variational expression.

Define $\alpha = \|e_1+e_2\|_X$ and $\beta = \|e_1-e_2\|_X$, and let $\phi$, $\psi \in B_{X^*}$ such that $\phi(e_1+e_2)=\alpha$ and $\psi(e_1-e_2)=\beta$. 
Write $\phi=(\phi_1,\phi_2)$ and $\psi=(\psi_1,\psi_2)$, so that $\alpha = \phi_1+\phi_2$ and $\beta = \psi_1 - \psi_2$. We compute
\[ 1 \geq \det (\psi,\phi) = \phi_2 \psi_1 - \phi_1 \psi_2 \geq \alpha + \beta -2.\]
To derive the last inequality, note that
\[ \phi_2 \psi_1 - \phi_1 \psi_2 - ( \alpha + \beta - 2) = (1 - \phi_2)(1-\psi_1) + (1-\phi_1)(1+\psi_2)\]
is nonnegative since $|\phi_i| \leq 1$ and $|\psi_j| \leq 1$.
We proved that $\alpha + \beta \leq 3$, and therefore either $\alpha \leq 3/2$ or $\beta \leq 3/2$. In the first case the conclusion is immediate; in the second case it suffices to
replace $e_2$ by $-e_2$.
\end{proof}

\begin{proof}[Proof of Theorem~\ref{theorem: 19/18}]
Apply Lemma~\ref{lemma:hexagon-dual} to both $X$ and $Y$, and consider the tensor
\[ z = 5 e_1 \otimes e_1 +  5 e_1 \otimes e_2 + 5 e_2 \otimes e_1 - 4 e_2 \otimes e_2 \in X \otimes Y .\] 
Consider also
\[ w^* = e_1^* \otimes e_1^* +  e_1^* \otimes e_2^* + e_2^* \otimes e_1^* - e_2^* \otimes e_2^* \in X^* \otimes Y^* .\]
Since the linear forms $e_i^*$ are bounded in absolute value by $1$ on the unit ball, an argument following closely the proof of the CHSH inequality~\cite{CHSH} shows that $\|w^*\|_{X^* \otimes_\e Y^*} \leq 2$. Together with the fact that $w^*(z)=19$, this implies that
\begin{equation} 
\label{eq:19/18lower} \|z\|_{X \otimes_{\pi} Y} \geq \frac{19}{2}.
\end{equation}
It remains to upper bound $\|z\|_{X \otimes_{\e} Y}$. Given $\phi \in B_{X^*}$ and $\psi \in B_{Y^*}$, consider the numbers 
\[ a = \phi(e_1), \ b = \phi(e_2), \ c = \psi(e_1), \ d = \psi(e_2) .\]
Both pairs $(a,b)$ and $(c,d)$ belong to the hexagon
\[ H = \left\{ (x,y) \in \R^2  : \ |x| \leq 1,\, |y| \leq 1,\, |x+y| \leq 3/2 \right\} .\]
Under these constraints it can be proved that
\[ 5ac+5ad+5bc-4bd \leq 9\, .\]
Indeed, it suffices to verify this inequality when $(a,b)$ and $(c,d)$ are extreme points of $H$; this yields a total of 36 different combinations to check. Finally, we have
\begin{equation}
\label{eq:19/18upper} \|z\|_{X \otimes_{\e} Y} = \sup_{\phi \in B_{X^*} ,\, \psi \in B_{Y^*}} (\phi \otimes \psi)(z) \leq 9\, . 
\end{equation}
Combining~\eqref{eq:19/18lower} with~\eqref{eq:19/18upper} gives $\|z\|_{X \otimes_\pi Y} \geq \frac{19}{18} \|z\|_{X \otimes_\e Y}$, as needed.
\end{proof}

%\begin{lemma} \label{lemma:octagons}
%Let $n \geq 2$. Any $n$-dimensional normed space is isometric to $(\R^n,\|\cdot\|)$, where $\|\cdot\|$ is a norm satisfying 
%\begin{equation} \label{eq:hexagon} \max \left( \|x\|_{\iy}, \frac{2}{3} |x_1+x_2|\right) \leq \| x \| \leq \|x\|_1 .\end{equation}
%for every $x=(x_1,\dots,x_n) \in \R^n$.
%\end{lemma}

\begin{remark}
The proof of Lemma~\ref{lemma:hexagon-dual} gives more information, namely that
\[ \|e_1+e_2\|_X \leq \alpha,  \ \|e_1 - e_2 \|_X \leq \beta \]
for some real numbers $\alpha$, $\beta \in [1,2]$ such that $\alpha + \beta \leq 3$. This extra information can presumably be used to improve the lower bound in Theorem~\ref{theorem: 19/18} to $\|z\|_{X \otimes_\pi Y} \geq \frac{8}{7} \|z\|_{X \otimes_\e Y}$ for an appropriate choice of $z$ depending on $\alpha$, $\beta$. However, since our arguments for that would rely heavily on computer assistance (and since the bound $8/7$ is unlikely to be optimal), we do not present them.
\end{remark}

Before we move on, let us discuss the special case of $2$-dimensional spaces. Although we are not yet able to evaluate the two quantities $r(2,2)$ and $r_s(2)$ exactly, we can show that
\begin{align}
    \frac43 &< r(2,2)\leq \sqrt2\, , \label{bounds r(2,2)}\\
    \frac43 &< r_s(2) \leq\sqrt3\, . \label{bounds r_s(2)}
\end{align}
To these inequalities we have to add the obvious fact that $r(2,2)\leq r_s(2)$. To justify the lower bound in~\eqref{bounds r(2,2)} (and hence also that in~\eqref{bounds r_s(2)}), we observe that combining~\eqref{Diameter in dim 2} and~\eqref{eq: lower bound d} yields $\rho(X,Y)\geq 4/3$ for all $2$-dimensional spaces $X,Y$. Equality is possible iff $\dbm(X,Y)=3/2$, which happens iff the unit balls of $X$ and $Y$ are simultaneous linear images of a square and a regular hexagon. Without loss of generality, this is the same as saying that $X$ is isomorphic to $\ell_1^2$. By Proposition~\ref{proposition:ell1}, this ensures that $\rho(X,Y)=\rho(\ell_1^2,Y)\geq \sqrt2>4/3$. Hence, it must be the case that $r(2,2)>4/3$ strictly.

As we have already seen, the upper bound in~\eqref{bounds r(2,2)} can by found by evaluating $\rho(\ell_1^2,\ell_2^2)=\sqrt2$, which is a special case of~\eqref{rho ell1 ell2}. The upper bound in~\eqref{bounds r_s(2)}, instead, is obtained by using as ansatz a space whose unit ball is a cleverly chosen octagon~\cite[Appendix~D]{lamiatesi}.

\subsection{An important special case: quantum theory} \label{subsec QM}

In this section we will study the case where both parties are described by a quantum model. Before we start, let us expound some notation that we already partially introduced. We denote by $\Herm{k}$ the real vector space of $k\times k$ Hermitian matrices. By equipping it with a Schatten norm $\|\cdot\|_p$, defined by $\|z\|_p\coloneqq \left(\Tr|z|^p\right)^{1/p}$, we can make such space a Banach space, which we denote by $S_p^{k,\mathrm{sa}}$. In what follows, we will be interested in the two particular cases $p=1$ and $p=\infty$, whose corresponding norms are the trace norm and the operator norm, respectively. For simplicity, we will make the canonical identification $(\Tk)^* = S_\iy^{k,\mathrm{sa}}$. Accordingly, the action of $(\Tk)^*$ on $\Tk$ is given simply by the Hilbert--Schmidt inner product, i.e.\ $y(x) = \Tr[xy]$ for $x\in \Tk$ and $y\in S_\iy^{k,\mathrm{sa}}$. As for the tensor product, remember that $\Herm{n}\otimes \Herm{m}=\Herm{nm}$ canonically. We now proceed to prove Theorem~\ref{theorem: rho for S1}, hence determining the scaling of the function $\rho(\Tn, \Tm)$ with respect to $n$ and $m$. The proof of the Corollary~\ref{cor data hiding} appears at the end of the present section.

\begin{note}
From now on, in some of the proofs we will find it convenient to adopt Dirac's notation for vectors and functionals in (or acting on) $\R^n$ and $\mathbf{C}^n$. This will be done without further comments.
\end{note}

\vspace{2ex}
\begin{proof}[Proof of Theorem~\ref{theorem: rho for S1}]

In order to establish~\eqref{eq: rho S1 scaling}, we have to show the existence of two constants $c,C>0$ such that 
\begin{equation} \label{eq:rho-QM}
c \min\{n, m\}^{3/2} \leq \sup_{0\neq z\in \Herm{n} \otimes \Herm{m}} \frac{\|z\|_{\Tn\otimes_\pi \Tm}}{\|z\|_{\Tn\otimes_\e \Tm}} \leq C \min\{n, m\}^{3/2} .
\end{equation}
We break down the argument to prove~\eqref{eq:rho-QM} into two parts.

\medskip

\textit{Step 1: lower bound.}
We assume without loss of generality that $n \leq m$, and consider two Hilbert--Schmidt orthonormal bases $(x_i)_{1 \leq i \leq n^2}$ and $(y_j)_{1 \leq j \leq m^2}$ of $\Herm{n}$ and $\Herm{m}$, respectively. We form the random tensor
\begin{equation}  z = \sum_{i=1}^{n^2} \sum_{j=1}^{m^2} g_{ij} x_i \otimes y_j, 
\end{equation}
where $(g_{ij})$ are independent $N(0,1)$ Gaussian random variables. Let us observe that the distribution of $z$ does not depend on the choice of the local orthonormal bases. We use the results from Corollary~\ref{chevet-schatten}: 
\begin{equation} \label{eq:1-epsilon-1}
\mathbb \E\|z\|_{\Tn \otimes_{\e} \Tm}\leq C \sqrt{n}\,m^{3/2},
\end{equation}
\begin{equation} \label{eq:inf-epsilon-inf}
\E\|z\|_{S_{\iy}^{n,\mathrm{sa} } \otimes_{\e} S_{\iy}^{m, \mathrm{sa}}} \leq C' \sqrt{m}.
\end{equation}
By duality,~\eqref{eq:inf-epsilon-inf} implies a lower bound on the projective norm of $z$ in $\Tn \otimes_\pi \Tm$. More precisely, using the duality between $\Tn \otimes_{\pi} \Tm$ and $S_{\iy}^{n,\mathrm{sa}} \otimes_{\e} S_{\iy}^{m,\mathrm{sa}}$ together with the Cauchy--Schwartz inequality, we obtain
\begin{equation} \label{eq:dualnorm} \E \left( \sum_{i,j} g_{ij}^2 \right)^{1/2} = \E \sqrt{\Tr[z^2]}  \leq   \sqrt{\E \| z\|_{\Tn \otimes_{\pi} \Tm}} \sqrt{ \E \| z\|_{S_{\iy}^{n,\mathrm{sa}} \otimes_{\e} S_{\iy}^{m,\mathrm{sa}}}}. \end{equation}
Since the l.h.s.\ of~\eqref{eq:dualnorm} is of order $nm$, combining~\eqref{eq:dualnorm} and~\eqref{eq:inf-epsilon-inf} yields the lower bound
\begin{equation} \label{eq:1-pi-1}
\mathbb \E\|z\|_{\Tn \otimes_{\pi} \Tm} \geq c n^2 m^{3/2}.
\end{equation}
Using the above relation together with~\eqref{eq:1-epsilon-1}, we see that the random variable 
\[ U \coloneqq C \|z\|_{\Tn \otimes_{\pi} \Tm} - cn^{3/2} \|z\|_{\Tn \otimes_{\e} \Tm} \] has a nonnegative expectation. In particular, the event $\{U \geq 0 \}$ is nonempty, from which it follows that $\rho(\Tn,\Tm) \geq cC^{-1} n^{3/2}$.

\medskip

\textit{Step 2: upper bound.}
As before, we assume, without loss of generality, that $n\leq m$. Let us consider an element $z\in \Herm{n} \otimes \Herm{m}$ such that $\|z\|_{\Tn\otimes_{\varepsilon} \Tm}=1$. By Corollary~\ref{cor state}, there is a state $\varphi$ such that
\begin{equation} \label{eq:upperbound-grothendieck}
\| \op{z}(x) \|_1 \leq 2 \sqrt{2} \left(\varphi(x^2) \right)^{1/2}
\end{equation}
for every $x \in \Herm{n}$ (the notation $\op{z}$ was introduced in~\eqref{tensor-operator}). In Dirac's notation, the spectral decomposition of $\varphi$ reads
\bbb
\varphi=\sum_{j=1}^n\lambda_j \ketbra{u_j}{u_j}\, ,
\eee
where $(\ket{u_j})_j$ is an orthonormal basis of $\mathbf{C}^n$, and $(\lambda_j)_j$ is a probability distribution. Then, it is clear that $E_{jk}\coloneqq \ketbra{u_j}{u_k}$, with $j,k=1,\cdots, n$, defines a Hilbert--Schmidt orthonormal basis of the space of $n\times n$ complex matrices. Using that $(E_{jk})_{jk}$, $(E_{kj})_{jk}$ is a biorthogonal system, we can write
\bbb
z=\sum_{j,k=1}^n E_{kj}\otimes \op{z}(E_{jk})\in S_1^n\otimes S_1^m\, .
\eee

If we define the Hermitian matrices $F_{jk}\coloneqq E_{jk}+E_{kj}$ and $H_{jk}\coloneqq i(E_{jk}-E_{kj})$, one can easily check that 
\bb
z=\sum_j E_{jj}\otimes \op{z}(E_{jj})+\frac{1}{2}\sum_{j < k}\left(F_{jk}\otimes \op{z}(F_{jk})+H_{jk}\otimes \op{z}(H_{jk})\right) ,
\label{eq: z hermitian decomposition}
\ee
where all indices range from $1$ to $n$. We then obtain the following:
\begin{align*}
    \|z\|_{\Tn\otimes_\pi \Tm} &\textleq{1} \sum_j \|E_{jj}\|_{1}\|\op{z}(E_{jj})\|_{1} +\frac{1}{2}\sum_{j<k} \left(\|F_{jk}\|_{1}\|\op{z}(F_{jk})\|_{1}+\|H_{jk}\|_{1}\| \op{z}(H_{jk})\|_{1}\right) \\
    &\textleq{2} \sum_j \|\op{z}(E_{jj})\|_{1} + \sum_{j<k} \left(\|\op{z}(F_{jk})\|_{1}+\| \op{z}(H_{jk})\|_{1}\right) \\
    &\textleq{3} 2\sqrt2 \sum_j \sqrt{\varphi(E_{jj})^2} + 2\sqrt2 \sum_{j<k} \left( \sqrt{\varphi(F_{jk}^2)}+\sqrt{\varphi(H_{jk}^2)}\right) \\
    &\texteq{4} 2\sqrt2 \sum_j \sqrt{\lambda_j} + 4\sqrt2 \sum_{j<k} \sqrt{\lambda_j+\lambda_k} \\
    &\textleq{5} 2\sqrt2 \left(\sqrt{n}\left(\sum\nolimits_j \lambda_j\right)^{1/2} + 2\sqrt{\frac{n(n-1)}{2}} \left(\sum\nolimits_{j < k} (\lambda_j +\lambda_k) \right)^{1/2} \right) \\
    &\texteq{6} 2\sqrt2 \left( \sqrt{n} + 2\sqrt{\frac{n(n-1)}{2}} \sqrt{n-1}\right) \\
    &= 4 n^{3/2} - 2\sqrt2 (\sqrt2 -1) \sqrt{n}\, .
\end{align*}
The justification of the above steps is as follows: 1: we used the decomposition~\eqref{eq: z hermitian decomposition} as an ansatz into the minimisation that defines the projective norm~\eqref{projective}; 2: we observed that $\|E_{jj}\|_1=1$ and $\|F_{jk}\|_1=2=\|H_{jk}\|_1$ for all $j<k$; 3: follows from~\eqref{eq:upperbound-grothendieck}; 4: we evaluated $\varphi(E_{jj}^2) = \varphi(E_{jj})=\lambda_j$ and 
\bbb
\varphi(F_{jk}^2) = \varphi(H_{jk}^2) = \varphi(|u_j\rangle\!\langle u_j| + |u_k\rangle\!\langle u_k|)= \lambda_j +\lambda_k\, ;
\eee
5: is an application of the Cauchy--Schwartz inequality; 6: we computed 
\bbb
\sum_{j<k} (\lambda_j +\lambda_k) = \frac12 \sum_{j\neq k} (\lambda_j+\lambda_k) = \frac12 \left(\sum_j (n-1) \lambda_j + \sum_k (n-1) \lambda_k\right) = n-1\, ,
\eee
and remembered that $\sum_j \lambda_j =1$. This completes the proof of~\eqref{eq: rho S1 upper precise}, which in turn implies~\eqref{eq: rho S1 scaling} with $C=4$.
\end{proof}

%Now, according to the definition of the projective tensor norm, we write  
%\begin{align}\label{split z} \|z\|_{\Tn\otimes_\pi \Tm}&\leq \sum_j \|E_{jj}\|_{1}\|\op{z}(E_{jj})\|_{1} +\frac{1}{2}\sum_{j<k}\left(\|F_{jk}\|_{1}\|\op{z}(F_{jk})\|_{1}+\|H_{jk}\|_{1}\| \op{z}(H_{jk})\|_{1}\right). \end{align}
%Now, while it is obvious that $\|E_{jj}\|_{1}=1$ and $\varphi(E_{jj}^2)=\lambda_j$ for all $j$, simple calculations show that for every pair $j,k$ such that $j\neq k$ we have that $\|F_{jk}\|_{1}=\|H_{jk}\|_{1}=2$ and
%\bbb \varphi(F_{jk}^2)=\varphi(H_{jk}^2)=\lambda_j+\lambda_k . \eee
%Hence, by combining~\eqref{split z} and~\eqref{eq:upperbound-grothendieck}, we infer that 
%\begin{align*} \|z\|_{\Tn\otimes_\pi \Tm} &\leq 2\sqrt{2}\left(\sum_j \sqrt{\lambda_j}+ \sum_{j < k}\sqrt{\lambda_j+\lambda_k}\right)\\ &\leq  2\sqrt{2} \left(\sqrt{n}\left(\sum\nolimits_j \lambda_j\right)^{1/2} + \sqrt{\frac{n(n-1)}{2}} \left(\sum\nolimits_{j < k} (\lambda_j +\lambda_k) \right)^{1/2} \right)\\ &= 2 \left( n^{3/2} + (\sqrt2 - 1) \sqrt{n} \right) , \end{align*}
%where the second step is an application of the Cauchy--Schwartz inequality. Upon taking the supremum over $z$, this implies that~\eqref{eq:rho-QM} holds with $C=2$ (asymptotically).

\begin{proof}[Proof of Corollary~\ref{cor data hiding}]
The claim~\eqref{eq: LO norm QM} derives from~\eqref{eq:rho-QM}. Indeed: (i) the norm $\|\cdot\|_{\mathrm{LO}}$ as defined by~\cite{VV-dh} satisfies $\|\cdot\|_{\mathrm{LO}}\leq \|\cdot\|_1$ e.g.\ by Helstrom's theorem; (ii) the inequality $\|\cdot\|_{1}\leq \| \cdot\|_{\Tn \otimes_\pi \Tm}$ follows from~\eqref{pi as base norm on AB} combined with the fact that the standard quantum mechanical composition rule yields a legitimate composite in the GPT sense; and (iii) $\|\cdot\|_{\mathrm{LO}}\geq \|\cdot\|_{\Tn\otimes_\e \Tm}$ by~\cite[Proposition~22]{ultimate}. 
\end{proof}

\section{Asymptotic lower bounds on the \texorpdfstring{$\pi/\e$}{pi/epsilon} ratio}

\label{sec asymptotic bounds}

\subsection{A lower bound for two copies of the same theory}

In this section we prove that  $r_s(n)\geq c\sqrt{n}/\log^3 n$ for a certain universal constant $c > 0$, which is the technically challenging part of the statement of Theorem~\ref{theorem: lower bound rs(n)}. In fact, remember that the example of $\ell_1^n$ shows that $r_s(n)\leq \sqrt{2n}$ (as reported in~\eqref{elementary ineq n=m}), hence the aforementioned result is optimal up to logarithmic factors. For an $n$-dimensional Banach space we denote $d_X=\dbm(X,\ell_2^n)$, where $\dbm$ is the Banach--Mazur distance.

\begin{theorem} \label{theorem: lower bound rho(X,X)}
There exists a universal constant $c>0$ such that for every Banach space $X$ of dimension $n$, we have 
\[
\rho(X,X)\geq \frac{cn}{d_X\log^3 n}\, .
\label{lower bound rho(X,X)}
\]
In particular, $r_s(n)\geq c\sqrt{n}/\log^3 n$.
\end{theorem}

\begin{proof}
The lower bound on $r_s(n)$ follows immediately by combining~\eqref{lower bound rho(X,X)} and the well-known estimate $d_X \leq \sqrt{n}$ in~\eqref{John estimate}. We now set out to prove~\eqref{lower bound rho(X,X)}. We may assume that $X$ is equal to $(\R^n, \| \cdot \|_X)$, with  
\bb
\|\id:\ell_2^n\rightarrow X\| \cdot \|\id:X\rightarrow \ell_2^n\| = d_X\, .
\label{BM distance ell 2}
\ee
Our main tool is the following lemma, whose proof we postpone. It is based on ideas from~\cite{BM86} (see also~\cite[Lemma~46.2]{TOMCZAK} and comments below it).

\begin{lemma}\label{Block-lemma}
Consider a Banach space $X = (\R^n, \|\cdot\|_X)$, and assume that $\frac{1}{\sqrt{n}} |\cdot| \leq \|\cdot\|_X \leq |\cdot|$, where $|\cdot|$ is the standard Euclidean norm. Then there exist orthonormal vectors $\ket{f_i} \in \R^n$, $i=1,\cdots, k$, with $k\geq cn/\log(n)$, such that
\bb
\E \left\|\sum_{i=1}^k g_i \ket{f_i} \right\|_X \E \left\|\sum_{i=1}^kg_i \bra{f_i} \right\|_{X^*}\leq Cn \log n\, .
\label{Block-lemma eq}
\ee
Here, $c$ and $C$ are universal constants and $(g_i)_{i=1}^n$ is a sequence of independent $N(0,1)$ Gaussian random variables.% and $d_X=\dbm (X, \ell_2^n)$ is the Banach--Mazur distance between $X$ and $\ell_2^n$.
\end{lemma}

%Note that
%\begin{align*}
%\ell^*(\tilde{T}^{-1})=\sup\{tr(\tilde{T}^{-1} S):\ell(S:\ell_2^n\rightarrow X)\leq 1\}=\sup\{tr(UP^{-1}U^*S):\ell(S:\ell_2^n\rightarrow X)\leq 1\}\\=\sup\{tr(P^{-1}U^*SU):\ell(S:\ell_2^n\rightarrow X)\leq 1\}=\sup\{tr(P^{-1}U^*\tilde{S}):\ell(\tilde{S}:\ell_2^n\rightarrow X)\leq 1\}\\=\ell^*(T^{-1})
%\end{align*}

%Since $\epsilon$ and $\pi$ norms are dual to each other, we have that for every $u\in X\otimes X$ and  $v\in X^*\otimes X^*$, $$\langle u, v\rangle\leq \|u\|_{X\otimes\pi X}\|v\|_{X^*\otimes\epsilon X^*}\leq Q_X \|u\|_{X\otimes\epsilon X}\|uv\|_{X^*\otimes\epsilon X^*}.$$Hence, we have 
%\begin{align*}
%Q_X\geq \sup_{\substack{u\in X\otimes X\\  v\in X^*\otimes X^*}}\frac{\langle u, v\rangle}{\|u\|_{X\otimes\epsilon X}\|uv\|_{X^*\otimes\epsilon X^*}}.
%\end{align*}

Let us consider the vectors $(\ket{f_i})_{1 \leq i \leq k} \in \R^n$ from Lemma~\ref{Block-lemma}, and form the random tensors $\ket{z}=\sum_{i,j=1}^kg_{ij} \ket{f_i} \otimes \ket{f_j}\in X\otimes X$ and $\bra{z}=\sum_{i,j=1}^kg_{ij} \bra{f_i}\otimes \bra{f_j}\in X^*\otimes X^*$, where $(g_{ij})$ are independent $N(0,1)$ Gaussian random variables. It is clear that 
\bb
\E \braket{z|z} = \E\sum_{i,j=1}^k g_{ij}^2=k^2\geq \frac{c^2 n^2}{\log^2 n}\, .
\label{tr zz lower estimate}
\ee
On the other hand, according to Chevet's inequality (Theorem~\ref{Chevet}), we have 
\begin{align}
\E \left\|\ket{z} \right\|_{X\otimes_\e X} &= \E\left\|\sum_{i,j=1}^kg_{ij}\ket{f_i} \otimes \ket{f_j} \right\|_{X\otimes_\e X}\leq 2\E\left\|\sum_{i=1}^kg_i\ket{f_i} \right\|_X\| \id_{K}:\ell_2^k\rightarrow X\|, \label{estimate Chevet primal} \\
\E \left\| \bra{z} \right\|_{X^*\otimes_\e X^*} &= \E\left\|\sum_{i,j=1}^kg_{ij}\langle f_i|\otimes \langle f_j|\right\|_{X^*\otimes_\e X^*}\leq 2\E \left\|\sum_{i=1}^kg_i\langle f_i|\right\|_{X^*}\|\id_{K}:\ell_2^k\rightarrow X^*\|, \label{estimate Chevet dual}
\end{align}
where $\id_K$ denotes the identity map restricted to $K\coloneqq \text{span}\{\ket{f_i} : \, 1 \leq i \leq k\}$. We can then write
\allowdisplaybreaks
\begin{align*}
\E \braket{z|z} &\textleq{1} \E\left\|\sum_{i,j=1}^k g_{ij} \ket{f_i} \otimes \ket{f_j}\right\|_{X\otimes_\pi X}\left \|\sum_{i,j=1}^k g_{ij} \bra{f_i}\otimes \bra{f_j} \right\|_{X^*\otimes_\e X^*} \\[.8ex]
&\textleq{2} \rho(X,X) \E\left\|\sum_{i,j=1}^k g_{ij} \ket{f_i} \otimes \ket{f_j}\right\|_{X\otimes_\e X}\left \|\sum_{i,j=1}^k g_{ij} \bra{f_i}\otimes \bra{f_j} \right\|_{X^*\otimes_\e X^*} \\[.8ex]
&\textleq{3} \rho(X,X)\left(\E\left\|\sum_{i,j=1}^kg_{ij}\ket{f_i} \otimes \ket{f_j}\right\|^2_{X\otimes_\e X}\right)^{\frac{1}{2}} \left(\E \left\|\sum_{i,j=1}^kg_{ij}\bra{ f_i}\otimes \bra{ f_j}\right \|^2_{X^*\otimes_\e X^*}\right)^{\frac{1}{2}}\\[.8ex]
&\textleq{4} C_2^2 \rho(X,X)\  \E\left\|\sum_{i,j=1}^kg_{ij}\ket{f_i} \otimes \ket{f_j}\right\|_{X\otimes_\e X}\E \left\|\sum_{i,j=1}^kg_{ij} \bra{f_i}\otimes \bra{f_j}\right \|_{X^*\otimes_\e X^*} \\[.8ex]
&\textleq{5} 4C_2^2 \rho(X,X)\ \E\left\|\sum_{i=1}^k g_i \ket{f_i} \right\|_X \E\left\|\sum_{i=1}^k g_i \bra{f_i}\right\|_{X^*} \|\id_{K}:\ell_2^k\rightarrow X\|\cdot\|\id_{K}:\ell_2^k\rightarrow X^*\| \\[.8ex]
&\textleq{6} 4 C_2^2 \rho(X,X) d_X\  \E\left\|\sum_{i=1}^k g_i \ket{f_i} \right\|_X \E\left\|\sum_{i=1}^k g_i \bra{f_i} \right\|_{X^*} \\[.8ex]
&\textleq{7} C' \rho(X,X) d_X\ n \log n\, .
\end{align*}
The justification of the above steps is as follows: 1: we used the duality of injective and projective norm~\eqref{duality epsilon pi}; 2: follows by definition of $\rho(X,X)$; 3: is an application of the Cauchy--Schwarz inequality; 4: is the $p=2$ case of the Khintchine--Kahane inequality (Theorem~\ref{Khinchin-Kahane}); 5: derives from~\eqref{estimate Chevet primal} and~\eqref{estimate Chevet dual}; 6: can be derived from~\eqref{BM distance ell 2}, using the fact that orthogonal projections onto subspaces of Hilbert spaces have norm $1$; 7: is the statement of Lemma~\ref{Block-lemma}. Combining the above estimate with the lower bound in~\eqref{tr zz lower estimate}, we deduce that
\bbb
\rho(X,X)\geq c\, \frac{n}{d_X \log^3 n}\, ,
\eee
which concludes the proof.
\end{proof}

\begin{proof}[Proof of Lemma~\ref{Block-lemma}]
%Since $\mathrm{d}(X,\ell_2^n) \leq \sqrt{n}$, we may assume that $\|\id:\ell_2^n\rightarrow X\|= 1$ and $\|\id:X\rightarrow \ell_2^n\| \leq \sqrt{n}$. %= d_X$. \tcr{This is not exactly John's position ; in general John's position gives a factor strictly larger than $d_X$}
%Note that in general, we know that there exists an isomorphism $T:\ell_2^n\rightarrow X$ such that $T(B_{\ell_2^n})\subset B_X\subset \sqrt{n} T(B_{\ell_2^n})$. We can then consider $B_Y=T^{-1}(B_X)$ and have $B_{\ell_2^n}\subset B_Y\subset \sqrt{n} B_{\ell_2^n}$. Then, since $X$ and $Y$ are isometric (by $T$), we can work equivalently with $Y$.
According to the $MM^*$-estimate (Theorem~\ref{ell-K}), there exists an isomorphism $T:\ell_2^n\rightarrow X$ such that
\begin{align}\label{ell-estimate lemma}
\ell_X(T)\ell_{X^*}((T^{-1})^*)\leq Cn \log n.
\end{align}
Moreover, since $\ell_X(T)=\ell_X(T\circ U)$ for every unitary $U$, it can be assumed that $T$ is positive definite. By the spectral theorem, $T$ can be written as
\bbb
T=\sum_{i=1}^n\lambda_i \ketbra{f_i}{f_i}\, ,
\eee
for some positive numbers $\lambda_i$ and $(|f_i\rangle)_{i=1}^n$ an orthonormal basis of $\R^n$.
Then, inequality~\eqref{ell-estimate lemma} implies that
\begin{align*}
\E\left\|\sum_{i=1}^n\lambda_ig_i |f_i\rangle\right\|_X \E \left\|\sum_{i=1}^n\lambda_i^{-1}g_i \langle f_i|\right\|_{X^*}
%&\leq  \left(\E\left\|\sum_{i=1}^n\lambda_ig_i |f_i\rangle\right\|^2_X\right)^{\frac{1}{2}}\left(\E \left\|\sum_{i=1}^n\lambda_i^{-1}g_i \langle f_i|\right\|^2_{X^*}\right)^{\frac{1}{2}}\\&
\leq Cn \log n\, .
\end{align*}
Using the inequalities $\| | f_i \rangle \|_X \geq 1/\sqrt{n}$ and $\| \langle f_j | \|_{X^*} \geq 1$ together with Jensen
inequality (or Lemma~\ref{Gaussian-Lemma}), we see that $\lambda_i\lambda_j^{-1} \leq C n^{3/2} \log n$ holds for any indices $i$, $j$. Let us denote $m=\min \{\lambda_k \, : \, 1 \leq k \leq n\}$ and $M=\max \{\lambda_k , \: \, 1 \leq  k \leq n\}$.  It follows that
\begin{align}\label{cardinal}
\frac{M}{m}\leq Cn^{3/2} \log n\, .
\end{align}
Now,~\eqref{cardinal} implies that the sets
\bbb
A_s = \left\{1\leq j\leq n \, : \, 2^{s-1}\leq \frac{\lambda_j}{m}\leq 2^s \right\}
\eee
with $s=1,\cdots, r$, define a partition of $\{1,\cdots, n\}$ for a certain $r\leq C'\log n$. By the pigeonhole principle, one can immediately deduce the existence of a set $A_{s_0}$ such that $|A_{s_0}|\geq \frac{n}{C'\log n}$. Now, consider the set of orthonormal vectors $\{|f_j\rangle\text{:   }j\in A_{s_0}\}$. Applying Lemma~\ref{Gaussian-Lemma}, we see that
\begin{align*}
\E \left\|\sum_{i\in A_{s_0}}g_i | f_i\rangle\right\|_X \E \left\|\sum_{i\in A_{s_0}}g_i \langle f_i|\right\|_{X^*}&
\leq \E \left\|\sum_{i\in A_{s_0}}\frac{\lambda_i}{2^{s_0-1}m }g_i |f_i\rangle \right\|_X \E\left\|\sum_{i\in A_{s_0}}2^{s_0} m \lambda_i^{-1}g_i  \langle f_i|\right\|_{X^*}\\ &\leq 2 \E\left\|\sum_{i\in A_{s_0}}\lambda_ig_i |f_i\rangle\right\|_X\E\left\|\sum_{i\in A_{s_0}}\lambda_i^{-1}g_i \langle f_i|\right\|_{X^*} \\
&\leq Cn \log n\, ,
\end{align*}
completing the argument.
\end{proof}

%\begin{remark}[\textbf{To be removed}]
%I don't see how to obtain $\rho(X,X)\geq C\frac{n}{d_X(1+\log d_X)^3}$. I guess this is not very important and maybe the statement should be $\rho(X,X)\geq C\frac{\sqrt{n}}{(1+\log n)^3}$ in any case. 
%My problem to get $(1+\log d_X)^3$ is that I don't see how to remove the $n$-dependence of the cardinal $r$ in Lemma~\ref{Block-lemma}
%\end{remark}

\subsection{A lower bound for any pair of theories}

The aim of this section is to prove Theorems~\ref{theorem: lower bound r(n,n)} and~\ref{theorem: lower bound r(n,m)}, which provide general lower bounds on $\rho(X,Y)$ as functions of the dimensions $n,m$ only. As discussed in Section~\ref{sec main results}, our strategy requires two preliminary results that allow to reduce the problem to the more manageable special case where either $X$ or $Y$ is one of the `classical' spaces $\ell_1^n$, $\ell_2^n$ or $\ell_{\iy}^n$. We start by presenting the solution to these special cases.

\begin{lemma} \label{lemma:lowerbound-ellp}
For every finite-dimensional normed space $X$ with $\dim X \geq n$, we have
\begin{enumerate}[(a)]
    \item $\rho(\ell_1^n,X) \geq \sqrt{n/2},$
    \item $\rho(\ell_2^n,X) \geq \sqrt{n},$
    \item $\rho(\ell_{\iy}^n,X) \geq \sqrt{n/2}.$
\end{enumerate}
\end{lemma}

\begin{proof}
We already know from Proposition \ref{proposition:ell1-X} that $\rho(\ell_1^n,X) = \pi_1^{(n)}(\id_X)$. Moreover (this is especially clear from \eqref{ell-1-N-X}), we have $\rho(\ell_1^n,Y) \leq \rho(\ell_1^n,X)$ whenever $Y$ is a subspace of $X$. Consequently, it suffices to prove (a) in the case when $\dim X=n$. In that case, we argue that
\[ \pi_1^{(n)}(\id_X) \geq \pi_2^{(n)}(\id_X) \geq \frac{1}{\sqrt 2}
\pi_2(\id_X) \geq \sqrt{n/2}, \]
where we used points 1., 2.\ and 3. from Proposition \ref{proposition p-summing}.

Part (b) is a direct consequence of Proposition \ref{proposition:john-factorization} together with the formulation of $\rho(\ell_2^n,X)$ from Lemma \ref{lemma: rho and trace(uv)}.

Finally, (c) follows from (a) since $\rho(\ell_{\iy}^n,X) = \rho(\ell_1^n,X^*)$, cf\ \eqref{eq: rho dual and exchange}.
\end{proof}

%\begin{lemma} \label{lemma:lowerbound-ellp_old}
%There is a constant $c>0$ such that for every finite-dimensional normed space $X$ with $\dim X \geq n$, we have
%\[ \rho(\ell_{1}^n,X) \geq c \sqrt{n},  \ \ \ \rho(\ell_{2}^n,X) \geq c \sqrt{n}, \ \ \ \rho(\ell_{\iy}^n,X) \geq c \sqrt{n}. \] 
%\end{lemma}

%\begin{proof}
%We use the proportional Dvoretzky--Rogers factorisation (Theorem~\ref{theorem:bourgainszarek}) applied to $X^*$, which can be reinterpreted as a factorisation of the formal identity map $\id : \ell_{2}^m \to \ell_{\iy}^m$ as $\id = TS$ %(\tcb{it should be $\id = TS$, right?})
%with $ \| S : \ell_2^m \to X^*\| \leq D$ and  $\| T : X^* \to \ell_{\iy}^m \| \leq 1$. Denote by $J : \ell_{\iy}^m \to \ell_2^m$ the formal identity map; then it is elementary to check that $\|J\| = \sqrt{m}$. The choices $u=SJ$, $v=T$ in~\eqref{eq:reformulation-trace} give $\rho(\ell_{\iy}^m,X) \geq \sqrt{m}/D$, and the choices $u=S$, $v=JT$ give $\rho(\ell_{2}^m,X) \geq \sqrt{m}/D$. To conclude we use the facts that $\f (\ell_2^m,\ell_2^n) = \f (\ell_{\iy}^m,\ell_{\iy}^n) = 1$ and~\eqref{eq: rho heredity f}. The case of $\ell_1^n$ follows by duality.
%\end{proof}

\begin{remark}
In light of the discussion at the end of Subsection~\ref{subsec XOR} (see also the Remark after Proposition~\ref{proposition:ell1}), we see that Lemma~\ref{lemma:lowerbound-ellp} entails the following: the gap between local and global bias for XOR games played over any system $AB$ in which e.g.\ $A=\mathrm{Cl}_n$ is a classical theory (defined in \eqref{classical probability theory GPT}) is at least $\sqrt{n/2}$ whenever $\dim B \geq n$.
\end{remark}

%\begin{remark}
%Could Lemma~\ref{lemma:lowerbound-ellp} be true with $c=1$ ? Ludovico noticed that the lower bound $\rho(\ell_1^2,X) \geq \sqrt{2}$ follows from the fact that the Euclidean space has maximal modulus of uniform convexity.
%\end{remark}

The following result is a variant of the `$\ell_1$/$\ell_2$/$\ell_{\iy}$ trichotomy' which is based on ideas from Pisier~\cite{Pisier80}, Rudelson~\cite{Rudelson95}, and Szarek--Tomczak-Jaegerman~\cite{STJ09}.

\begin{theorem} \label{theorem: trichotomy}
Let $X$ be a normed space of dimension $n$. Then for every $1 \leq A \leq \sqrt{n}$ at least one of the following holds
\begin{enumerate}
 \item $X$ contains a subspace of dimension $d:=c \sqrt{n}$ which is $C A \sqrt{\log n}$-isomorphic to $\ell_{\iy}^d$.
 \item $X^*$ contains a subspace of dimension $d$ which is $C A \sqrt{\log n}$-isomorphic to $\ell_{\iy}^d$.
 \item $X$ contains a $C\log n$-complemented $4$-Euclidean subspace of dimension $c A^2 / \log n$.
\end{enumerate}
Here, $C$ and $c$ are universal constants.
\end{theorem}

\begin{proof}
By the $MM^*$-estimate (Theorem~\ref{ell-K}), we may assume that $X = (\R^n,\|\cdot\|_X)$ with
\[ \ell_X(\id) \leq C \sqrt{n \log n} \ \ \textnormal{and} \ \ \ell_{X^*}(\id) \leq C \sqrt{n \log n} .\]
Let $\mathcal{E}$ be the John ellipsoid of $X$ as defined in Theorem~\ref{theorem:john}, $(|e_i\rangle)_{1 \leq i \leq n}$ be the semiaxes of
$\mathcal{E}$ and $(\lambda_i)$ their lengths, i.e.\ $\mathcal{E} = T(B_2^n)$ where $T = \sum \lambda_i \ketbra{e_i}{e_i}$. Assume also that $\lambda_1 \leq \lambda_2 \leq \cdots \leq \lambda_n$. Note that we can assume that $T$ is of this form because $T\circ u$ defines the same ellipsoid for every unitary $u$. We consider the following dichotomy.

\medskip

{\bfseries Case (i) : $\lambda_{n/3} \leq A/\sqrt{n}$.} Let $E = \mathspan \{ |e_i \rangle \, : \, 1 \leq i \leq n/3 \}$ and $P_E$ be the orthogonal projection onto $E$. We note that $P_E$ is orthogonal for both the standard Euclidean structure in $\R^n$ and the Euclidean structure induced by $\mathcal{E}$ (i.e.\ using $(\lambda_i |e_i\rangle)$ as an orthonormal basis). We apply Theorem~\ref{theorem: vershynin} to $P_E$ in order to produce an $m$-dimensional subspace of $X$ which is $R$-isomorphic to $\ell_{\iy}^m$, for $m=c\sqrt{n}$ and $R = C \ell'_X(P_E)$, where we denote by $\ell'_X$ the $\ell_X$-norm computed using the Euclidean structure induced by $\mathcal{E}$. We use Lemma~\ref{Gaussian-Lemma} to obtain the bound
\[ \ell'_X(P_E) = \E \left\| \sum_{i=1}^{n/3} g_i \lambda_i | e_i \rangle \right\|_X \leq \lambda_{n/3} \ell_X(P_E) \leq \frac{A}{\sqrt{n}} \ell_X(\id) \leq C A \sqrt{\log n} ,\]
and we conclude that $X$ contains a subspace which is $C A \sqrt{\log n}$-isomorphic to $\ell_{\iy}^{c \sqrt{n}}$.

\medskip

{\bfseries Case (ii) : $\lambda_{n/3} > A/\sqrt{n}$.} Let $F = \mathspan \{ e_i \st i > n/3 \}$ and denote
by $\id_F : F \to \R^n$ the identity map restricted to $F$. We have
%\[ K \cap F \supset \mathcal{E} \cap F \supset \frac{A}{\sqrt{n}} B_2^n \cap F.\]
\[ \| \id_F : \ell_2^n \to X \| \leq \frac{\sqrt{n}}{A}. \]

To see the previous bound, just write $\id_F=(T\circ T^{-1}|_F)$ and use that $\|T^{-1}|_F:\ell_2^n\to \ell_2^n\|\leq \sqrt{n}/A$ and $\|T:\ell_2^n\to X\|\leq 1$.
\medskip

We apply the same dichotomy to $X^*$. If case (i) occurs for either $X$ or $X^*$, we are done. It remains to consider the situation when case (ii) occurs for both. This means that there exist subspaces $F_1$ and $F_2$ of dimension $2n/3$ such that 
\[ \| \id_{F_1} : \ell_2^n \to X \| \leq \sqrt{n}/A,  \]
\[ \| \id_{F_2} : \ell_2^n \to X^* \| \leq \sqrt{n}/A. \]
Consider the subspace $F= F_1 \cap F_2$ (note that $\dim F \geq n/3)$. We are going to apply the Dvoretzky--Milman theorem (Theorem~\ref{theorem: dvoretzky}) to both $X \cap F$ and $X^* \cap F$ (that is, to the space $F$ with the norms inherited from $X$ and from $X^*$ respectively). The corresponding Dvoretzky dimensions are
\[ k_*(X \cap F) = \left( \frac{\ell_{X \cap F}(\id_F)}{\| \id_F : \ell_2^n \to X \| } \right)^2 \geq  \frac{A^2}{n} \ell_{X \cap F} (\id_F)^2 = A^2\ell_X(P_F)^2/n , \]
\[ k_*(X^* \cap F) = \left( \frac{\ell_{X^* \cap F}(\id_F)}{\| \id_F : \ell_2^n \to X^* \|} \right)^2 \geq \frac{A^2}{n} \ell_{X^* \cap F}(\id_F) = A^2 \ell_{X^*}(P_F)^2/n.\]
On the other hand, if we consider the random vector $g = \sum_{i=1}^{\dim F} g_i f_i$, where $(g_i)$ are independent $N(0,1)$ Gaussian random variables and $(f_i)$ is an orthonormal basis in $F$, we compute
\[ \frac{n}{3} \leq \dim(F) = \E | g |^2 \textleq{1.} \left( \E \left\| g \right\|^2_X  \right)^{1/2} \left( \E \left\| g \right\|^2_{X^*} \right)^{1/2} \textleq{2.} C \E \|g\|_X \E \|g\|_{X^*} = C \ell_X(P_F) \ell_{X^*}(P_F), \] 
where 1.\ follows from Cauchy--Schwarz inequality together with the inequality $|g|^2 \leq \|g\|_X\|g\|_{X^*}$, and 2.\ from Khintchine--Kahane inequalities (Theorem~\ref{Khinchin-Kahane}).
Since $\ell_{X^*}(P_F) \leq \ell_{X^*}(\id) \leq C \sqrt{n \log n}$,
% (\tcb{it is probably better to say ``Since $\ell_{X^*}(P_F) \leq \ell_{X^*}(\id) \leq C \sqrt{n \log n}$'', because it is what were are using in that sentence, right? Although we will use both estimates...}), 
we have $\ell_X(P_F) \geq c \sqrt{n/\log n}$, and similarly $\ell_{X^*}(P_F) \geq c \sqrt{n/\log n}$. It follows that
\[ k_*(X \cap F) \geq c A^2 / \log n \ \ \textnormal{and} \ \ k_*(X^* \cap F) \geq c A^2 / \log n. \]
By Dvoretzky--Milman theorem (Theorem~\ref{theorem: dvoretzky}) and the remark following it, there is a subspace $E \subseteq F$ of dimension $c A^2 / \log n$ such that both $X \cap E$ and $X^* \cap E$ are $4$-Euclidean. Moreover, using the extra information given by Theorem~\ref{theorem: dvoretzky}, we have
\[ \| \id_E : \ell_2^n \to X \| \leq \frac{2 \ell_X(P_F)}{\sqrt{n/3}} ,\]
\[ \| \id_E : \ell_2^n \to X^* \| \leq \frac{2 \ell_{X^*}(P_F)}{\sqrt{n/3}} .\]
Since $\| P_E : X \to \ell_2^n \| = \|P_E : \ell_2^n \to X^* \| = \|\id_E : \ell_2^n \to X^* \|$, we have
\begin{align*} \| P_E : X \to X \| & \leq  \| P_E : X \to \ell_2^n \| \cdot \| P_E : \ell_2^n \to X \| \\
& \leq  \frac{C \ell_X(P_F) \ell_{X^*}(P_F)}{n} \\ 
& \leq  \frac{C \ell_X(\id) \ell_{X^*}(\id)}{n} \\
& \leq  C' \log n
\end{align*}
and therefore $E$ is a $C'\log n$-complemented $4$-Euclidean subspace of $X$.
\end{proof}

\begin{proof}[Proof of Theorem~\ref{theorem: lower bound r(n,n)}]
As usual, the difficult part is to establish the lower bound on $r(n,n)$, while the upper bound is reported in~\eqref{elementary ineq n,m} (set $n=m$). Let $X$ and $Y$ be $n$-dimensional normed spaces, and $A>1$ be a number whose value will be optimised later. Theorem~\ref{theorem: trichotomy} implies in particular that at least one of the following occurs (here we use the classical fact that a subspace isometric to $\ell_{\iy}^m$ is automatically $1$-complemented): 
\begin{enumerate}
\item[(i)] $\f \left(\ell_{\iy}^{c\sqrt{n}},X\right) \leq CA \sqrt{\log n}$;
\item[(ii)] $\f \left(\ell_{1}^{c\sqrt{n}},X\right) \leq CA \sqrt{\log n}$; or
\item[(iii)] $\f \left(\ell_{2}^{cA^2/\log n},X\right) \leq C \log n$.
\end{enumerate}
If (i) holds, then by Lemma~\ref{lemma:lowerbound-ellp} and~\eqref{eq: rho heredity f} we obtain
\bbb
\rho(X,Y) \geq \rho\left(\ell_{\iy}^{c\sqrt{n}},Y\right) \Big/ \f \left(\ell_{\iy}^{c\sqrt{n}},X\right) \geq \frac{cn^{1/4}}{A \sqrt{\log n}}\, .
\eee
A similar estimate applies when $X$ satisfies (ii), or when $Y$ satisfies (i) or (ii). The only uncovered case is when $X$ and $Y$ both satisfy
(iii), and we have then
\bbb
\rho(X,Y) \geq  \rho\left( \ell_{2}^{cA^2/\log n},\ell_{2}^{cA^2/\log n}\right) \Big/ (C \log n)^2 \geq \frac{c A^2}{\log^3 n}\, .
\eee
The optimal choice is $A = n^{1/12} (\log n)^{5/6}$, which gives the announced lower bound.
\end{proof}

\begin{proof}[Proof of Theorem~\ref{theorem: lower bound r(n,m)}]
The upper bound on $r(n,m)$ follows again from~\eqref{elementary ineq n,m}, so we focus on the lower bound.
Let $X$ and $Y$ be normed spaces of respective dimensions $n$ and $m$ (with $n \leq m$), and $A>1$ be a number whose value will be optimised later. As is the previous proof, we combine Theorem~\ref{theorem: trichotomy} (applied only to $X$) and Lemma~\ref{lemma:lowerbound-ellp}. In case (i), we have
\bbb
\rho(X,Y) \geq \rho\left(\ell_{\iy}^{c \sqrt{n}},Y\right)\Big/\f \left(\ell_{\iy}^{c \sqrt{n}},X\right) \geq \frac{cn^{1/4}}{A \sqrt{\log n}}\, .
\eee
Case (ii) is similar by duality. In case (iii), we have
\bbb
\rho(X,Y) \geq \rho\left(\ell_{2}^{cA^2/\log n},Y\right)\Big/\f \left(\ell_{2}^{cA^2/\log n},X\right) \geq \frac{cA}{(\log n)^{3/2}} .
\eee
The optimal choice $A = n^{1/8} \sqrt{\log n}$ always gives the lower bound $\rho(X,Y) \geq cn^{1/8}/\log n$, concluding the proof.
\end{proof}

\section{Conclusions}

In this paper, we have defined and investigated XOR games from the foundational standpoint provided by general probabilistic theories. This has led us to identify a deep connection between the minimum relative increase in the bias when global strategies displace local ones on the one hand, and the so-called projective/injective ratio on the other. The existence of such a connection is made possible by the fact that all norms on a given vector space can be well approximated by suitable base norms induced by GPTs.

The projective/injective ratio $r(n,m)$ is a universal function over pairs of integers that encodes some structural information about the theory of Banach spaces. For instance, we have shown that $n/r(n,n)$ provides a lower bound on the diameter of the Banach--Mazur compactum in dimension $n$ as measured by the weak distance~\cite{Tomczak84, Rudelson95}. We have also proved that $r(n,m)$ is always lower bounded by a universal constant strictly larger than $1$. This shows the remarkable fact that injective and projective tensor product can never be isometric, even though Pisier's celebrated construction~\cite{Pisier} demonstrates that they can be isomorphic in the case where the spaces have infinite dimension. Along the way, we developed an Auerbach--type lemma that may be of independent interest.

The main results we have presented concern the asymptotic behaviour of the ratio $r(n,m)$ and of its symmetrised version $r_s(n)$. In this context, we were able to show that, up to logarithmic factors, $r_s(n)$ is of the order $\sqrt{n}$. We showed that $r(n,m)$ grows at least as $\min\{n,m\}^{1/8}$, and that one can improve the exponent to $1/6$ if $n=m$. The proofs of these latter results follow by putting together an understanding of the projective/injective ratio in tensor products of the form $\ell_p^d\otimes X$, with $p=1,2,\infty$, and a `trichotomy theorem' that identifies in any normed space a sufficiently large subspace that is close in the Banach--Mazur distance to either (a) $\ell_1^d$, or (b) $\ell_2^d$, or (c) $\ell_\infty^d$. The main technical hurdle consists in establishing the additional requirement that in case (b) the chosen subspace is also well-complemented. As we have discussed, our findings draw on previous ideas by Pisier, Rudelson, Szarek, and Tomczak-Jaegermann.

Finally, although our primary subject of study is the intrinsic theory of XOR games played over general physical systems, it would be futile to deny that quantum systems hold great importance, due to their omnipresence in Nature as we currently understand it. In this spirit, we determined the exact scaling of the maximal global/local bias ratio in quantum XOR games, finding that it is of the order $\min\{n,m\}^{3/2}$, with $n,m$ being the local Hilbert space dimensions. Interestingly, this implies a new bound on the maximal strength of quantum mechanical data hiding against local operations.

Our work leaves a number of open problems that we believe are worth investigating. Let us recall briefly some of them. First, it would be interesting to compute exactly the absolute minimum of $r(n,m)$ across all pairs of integers, which one may conjecture to be equal to $\sqrt2$. A perhaps more profound question is to determine the best exponent $\gamma_{\text{opt}}$ such that $r(n,m)\geq c\min\{n,m\}^{\gamma_{\text{opt}}}$ for all $n,m$. We ask whether $\gamma_{\text{opt}}=1/2$. As we have seen, the simplified statement with $n=m$ would follow from Rudelson's conjecture~\cite{Rudelson95} that the Banach--Mazur compactum in dimension $n$ has a diameter of the order $\sqrt{n}$ with respect to the weak distance.

\vspace{3ex}
\emph{Acknowledgements.} We are grateful to Gilles Pisier for sharing with us an unpublished proof of the estimate $r_s(n)\geq n^{1/10 - o(1)}$, and to Marius Junge for helpful comments on some of our results. We thank the Institut Henri Poincar\'{e} for support and hospitality during the programme `Analysis in Quantum Information Theory' when part of the work on this paper was performed. GA was supported in part by ANR (France) under the grant StoQ (2014-CE25-0003). LL acknowledges financial support from the European Research Council (ERC) under the Starting Grant GQCOP (Grant no.~637352). CP is partially supported by the Spanish `Ram\'on y Cajal Programme' (RYC-2012-10449), the Spanish `Severo Ochoa Programme' for Centres of Excellence (SEV-2015-0554) and the grant MTM2014-54240-P, funded by Spanish MINECO. The research of SJS was supported in part by a grant DMS-1600124 from the National Science Foundation (U.S.A.). AW acknowledges support from the Spanish MINECO, project FIS2016-86681-P, with the support of FEDER funds, and the Generalitat de Catalunya, CIRIT project 2014-SGR-966.

\bibliography{biblio}

\appendix

\section{More on XOR games in GPTs} \label{app XOR games in GPTs}

Throughout this appendix we will demonstrate that the connection drawn by Theorem~\ref{theorem: bias xor games} between injective norms and bias of XOR games under local strategies is in a certain sense robust. Namely, we will show that allowing the players to use a bounded amount of back-and-forth communication does not make the bias larger that a constant times the same injective norm as in~\eqref{local bias}. In other words, the bias does not grow by more than a constant factor with respect to the purely local case.

Our argument does not require the communication to be classical. In fact, in principle the players are allowed to exchange any physical system described by a GPT. For instance, Alice could initiate the protocol by manipulating the subsystem $A$ corresponding to her share of the question so as to prepare a bipartite state of a new system $A_1M_1$; the subsystem $M_1$ is sent to Bob, while Alice keeps $A_1$ for later use; then, Bob employs $M_1$ together with his share of the question $B$ to prepare a message $M'_1$ to be sent to Alice and a record $B_1$ for later use. After $N$ such rounds, Alice will have sent the systems $M_1,\ldots, M_N$, and Bob will have sent the systems $M'_1,\ldots, M'_N$. The total dimension of the systems exchanged is thus
\bb
L_\leftrightarrow \coloneqq (\dim (M_1)\ldots \dim (M_N)) (\dim (M'_1)\ldots \dim (M'_N))\, .
\label{eq: total dimension messages}
\ee
In what follows, we will refer to such a setting as a `local strategy assisted by two-way communication of total dimension $L_\leftrightarrow$'. We now deal with the problem of bounding the corresponding bias.

\begin{note}
%Throughout the rest of this section, we will simplify the notation by identifying with the same capital letter a GPT and the vector space over which it is constructed. Accordingly, the algebraic tensor product of two systems is usually denoted by simple juxtaposition, e.g.\ $V_A\otimes V_B \mapsto AB$. If we want to specify a norm on the tensor product, we may write things like $A\otimes_\e B$ or $A\otimes_\pi B$ (remember that $A$ and $B$ are naturally equipped with their base norms).
We will often consider complicated compositions of maps acting on different systems. The convention we adopt is to omit all occurrences of the identity map acting on the untouched systems. In this way, if $T:A\to BC$ and $S:B\to DE$ are linear maps, we write $S T$ instead of $(S_B\otimes \id_C)\, T_A$.
\end{note}

%\begin{lemma} \label{lemma: pos-norm-pres maps constractions}
%Let $A,B,A',B'$ be four GPTs, and let the composite cones $C_{AB}$ and $C_{A'B'}$ satisfy the bounds~\eqref{CAB bound}. Consider a map $T:AB\to A'B'$ that is: %(i) positive, in the sense that $T(C_{AB})\subseteq C_{A'B'}$; (ii) normalisation-preserving, i.e.\ such that $T^*(u_{A'}\otimes u_{B'})=u_A\otimes u_B$. Then
%\bb \left\| T: A\otimes_\pi B\to A'\otimes_\e B'\right\| \leq 1\, . \label{eq: pos-norm-pres maps contractions} \ee
%\end{lemma}

%\begin{proof} By employing~\eqref{base norm}, it is not difficult to see that properties (i) and (ii) imply that $T$ is a contraction with respect to the local base norms $\|\cdot\|_{AB}$ and $\|\cdot\|_{A'B'}$. However, it follows from~\eqref{pi as base norm on AB} that $\|\cdot\|_{AB}\leq \|\cdot\|_{A\otimes_\pi B}$, while Theorem~\ref{theorem: bias xor games} entails among other things that $\|\cdot\|_{A'B'}\geq \|\cdot\|_{A'\otimes_\e B'}$. Hence, $T:A\otimes_\pi B\to A'\otimes_\e B'$ is a fortiori a contraction. \end{proof}

\begin{proposition} \label{proposition: bias xor games 2-way}
Let $G=(AB, \omega, p,c)$ be an XOR game over a bipartite GPT $AB$, and set $z_G = \sum_{i} p_i (-1)^{c_i}\omega_i$ as in Theorem~\ref{theorem: bias xor games}. The bias $\beta_{\leftrightarrow}(G)$ of $G$ under local strategies assisted by two-way communication of total dimension $L_\leftrightarrow$ can be upper bounded as
\bb
\beta_{\leftrightarrow}(G) \leq \sup_{\dim W\leq L_\leftrightarrow} \left\|\op{z}_G\otimes \id_W : V_A^* \otimes_\e W \longrightarrow V_B\otimes_\pi W \right\| \leq L_\leftrightarrow\, \|z_G\|_{V_A\otimes_\e V_B} = L_\leftrightarrow\, \beta_{\mathrm{LO}}(G)\, ,
\label{eq: bounded 2-way bias}
\ee
where the optimisation is over all normed spaces $W$ of dimension up to $L_\leftrightarrow$, and $\op{z}_G:V_A^*\to V_B$ is the linear map associated with the tensor $z_G\in V_A\otimes V_B$ according to~\eqref{tensor-operator}.
%\bb \begin{aligned} \|w\|_{\left(A\otimes_{\leftrightarrow} B\right)^*}  =\ &\mathrm{inf} & &\!\! \sum_{\beta=1}^m\ \prod_{\alpha=1}^{n+1} \left\|S_\alpha^\beta\!: B_{\alpha-1}\!\otimes_\pi\! M_{\alpha}\!\to\! B_\alpha\! \otimes_\e\! M'_\alpha\right\| \left\| T_\alpha^\beta\!: A_{\alpha\!-\!1}\!\otimes_\pi\! M'_{\alpha\!-\!1}\!\to\! A_\alpha\! \otimes_\e\! M_\alpha \right\| \\ &\mathrm{s.t.} & &\!\! \left\{T_\alpha^\beta: A_{\alpha-1} M_{\alpha-1}'\to A_\alpha M_\alpha \right\}_{\alpha=1,\ldots, n+1}^{\beta=1,\ldots, r} \\ & & &\!\! \left\{ S_\alpha^\beta: B_{\alpha-1}M_\alpha\to B_\alpha M'_\alpha\right\}_{\alpha=1,\ldots,n+1}^{\beta=1,\ldots, r} \\ & & & \op{w} = \sum_{\beta=1}^r (S^\beta_{n+1}\circ T^\beta_{n+1})\circ\ldots\circ (S^\beta_1\circ T^\beta_1)\, . \end{aligned} \ee
\end{proposition}

\begin{proof}
The $n$ rounds of communication can be represented by linear maps $T_\alpha: V_{A_{\alpha-1}} \otimes V_{M_{\alpha-1}'}\to V_{A_\alpha}\otimes V_{M_\alpha}$ and $S_\alpha: V_{B_{\alpha-1}}\otimes V_{M_\alpha} \to V_{B_\alpha}\otimes V_{M'_\alpha}$, for $\alpha=1,\ldots, N$, where for convenience we identified $A_0\coloneqq A$, $B_0\coloneqq B$, and $V_{M'_0}\coloneqq \R$. After the communication stage has taken place, Alice is left with the systems $A_N M'_N$, while Bob will have only $B_N$. They then perform local measurements to output the answers. These can be conveniently represented as $\left\{\frac{u+\varphi}{2}, \frac{u-\varphi}{2}\right\}$ (on Alice's side) and $\left\{\frac{u+\psi}{2}, \frac{u-\psi}{2}\right\}$ (on Bob's side), where $\varphi\in V_{A_N}^*\otimes V_{M'_N}^*$ and $\psi\in V_{B_N}^*$. A reasoning analogous to that in the proof of Theorem~\ref{theorem: bias xor games} shows that the bias for this strategy will be given by
\bb
\beta = (\varphi\otimes \psi)\left((S_N T_N)\ldots (S_1 T_1)(z_G)\right) \eqqcolon w(z_G) = \tr[\op{z}_G\, \op{w}\,^* ]\, ,
\label{eq: bias 2-way strategy}
\ee
where we observed that the validity of the above equation for all $z_G$ defines a functional $w\in V_{A}^*\otimes V_B^*$ (which depends on $\varphi$, $\psi$, and all the maps $T_\alpha, S_\alpha$, for $\alpha=1,\ldots, N$), and for the last step we used~\eqref{evaluation = trace}.

Now, we claim that the rank of the operator $\op{w}:V_A\to V_B^*$ satisfies
\bb
\rk\left(\op{w}\right) \leq L_\leftrightarrow\, .
\label{eq: rank w}
\ee
This can be verified straightforwardly by considering for all $\alpha$ families of vectors $\{x_{\alpha, j_\alpha}\in V_{M_\alpha}\}_{j_\alpha=1,\ldots, \dim(M_\alpha)}$, $\{y_{\alpha, k_\alpha}\in V_{M'_\alpha}\}_{k_\alpha=1,\ldots, \dim (M_\alpha')}$ and families of maps $\big\{\overline{T}_{\alpha, j_\alpha}:V_{A_{\alpha-1}} \otimes V_{M_{\alpha-1}'}\to V_{A_\alpha}\big\}_{j_\alpha=1,\ldots, \dim(M_\alpha)}$, $\big\{\overline{S}_{\alpha, k_\alpha}: V_{B_{\alpha-1}}\otimes V_{M_\alpha} \to V_{B_\alpha} \big\}_{k_\alpha=1,\ldots, \dim( M_\alpha')}$ such that one can expand
\bbb
T_\alpha = \sum_{j_\alpha=1}^{\dim(M_\alpha)} \overline{T}_{\alpha, j_\alpha} \otimes x_{\alpha, j_\alpha}\, ,\qquad S_\alpha = \sum_{k_\alpha=1}^{\dim(M_\alpha')} \overline{S}_{\alpha, k_\alpha} \otimes y_{\alpha, k_\alpha}\, .
\eee
Defining the `reduced' maps $\overline{T}_{\alpha, j_\alpha} y_{\alpha-1,k_{\alpha-1}}: V_{A_{\alpha-1}} \to V_{A_\alpha}$ (for $\alpha=2,\ldots, N$) and $\overline{S}_{\alpha, k_\alpha}x_{\alpha,j_{\alpha}}: V_{B_{\alpha-1}} \to V_{B_\alpha}$ (for $\alpha=1,\ldots, N$), we see that
\bbb
w = \sum_{\{j_\alpha,\, k_\alpha\}_\alpha} \!\!\varphi \left( y_{N,k_N}\otimes (\overline{T}_{N,j_N}y_{N-1,k_{N-1}})\!\ldots\!(\overline{T}_{2,j_2}y_{1,k_1}) \overline{T}_{1,j_1} \right)\,\otimes\, \psi\left( (\overline{S}_{N,k_N} x_{N,j_N})\!\ldots\! (\overline{S}_{1,k_1} x_{1,j_1}) \right) ,
\eee
where the first tensor factors are functional in $V_A^*$, and the second belong to $V_{B}^*$. Since the above sum contains exactly $L_\leftrightarrow$ terms, we see that~\eqref{eq: rank w} follows.

As it turns out, $w$ satisfies also the inequality
\bb
\|\op{w}:V_A\to V_B^*\| \leq 1\, .
\label{eq: injective norm w}
\ee
To see why this is the case, observe that the bias $\beta = w(z_G)$ cannot be larger than the maximal bias achievable by global strategies, as given by Theorem~\ref{theorem: bias xor games}. This implies that $w(z_G)\leq \|z_G\|_{V_A\otimes_\pi V_B}$. Since this has to hold for all $z_G\in V_A\otimes V_B$, and injective and projective tensor product are dual to each other by~\eqref{duality epsilon pi}, we deduce that $1\geq \|w\|_{(V_A\otimes_\pi V_B)^*}=\|w\|_{V_A^*\otimes_\e V_B^*}=\|\op{w}:V_A\to V_B^*\|$, where the last equality is an application of~\eqref{injective = operator norm}.

Putting together~\eqref{eq: bias 2-way strategy},~\eqref{eq: rank w}, and~\eqref{eq: injective norm w}, we see that
\bbb
\beta_{\leftrightarrow}(G) \leq \sup_{\substack{\\ w\in V_A^*\otimes V_B^* \\[.3ex] \rk(\op{w})\leq L_\leftrightarrow \\[.3ex] \|\op{w}:V_A\to V_B^*\|\leq 1}} |w(z)| = \sup_{\substack{\\ \|F:V_A\to V_B^*\|\leq 1 \\[.3ex] \rk(F)\leq L_\leftrightarrow}} \tr\left[ \op{z}_G\, F^*\right] .
\eee
As follows from elementary linear algebra, an operator $F:V_A\to V_B^*$ satisfies $\rk(F)\leq L_\leftrightarrow$ and $\|F\|\leq 1$ if and only if it can be factorised as $F = f_2 f_1$, where $f_1:V_A\to W$, $f_2:W\to V_B^*$ are linear maps, $W$ is a suitable Banach space of dimension $\dim W\leq L_\leftrightarrow$, and $\|f_1\|, \|f_2\|\leq 1$. Using this observation, we can rewrite the upper bound in the above inequality as
\bb
\beta_{\leftrightarrow}(G) \leq \sup_{\substack{\\ \dim W\leq L_\leftrightarrow \\[.3ex] \|f_1:V_A\to W\|,\, \|f_2:W\to V_B^*\| \leq 1}} \tr\left[ \op{z}_G\, f_1^* f_2^*\right] .
\label{eq: bias 2-way upper bound 1}
\ee
Defining the tensors $u\in V_A^*\otimes W$ and $v\in V_B^*\otimes W^*$ such that $\op{u}=f_1$ and $\op{v}=f_2^*$, we can rewrite
\bbb
\tr\left[ \op{z}_G\, f_1^* f_2^*\right] = v\left( (\op{z}_G\otimes \id_W)(u) \right).
\eee
At the same time, the constraints $\|f_1:V_A\to W\|, \|f_2:W\to V_B^*\|\leq 1$ become simply $\|u\|_{V_A^* \otimes_\e W},\, \|v\|_{V_B^*\otimes_\e W^*}\leq 1$. Using once again~\eqref{duality epsilon pi}, the bound in~\eqref{eq: bias 2-way upper bound 1} translates to
\begin{align*}
\beta_\leftrightarrow(G) &\leq \sup_{\substack{\\ \dim W\leq L_\leftrightarrow \\[.3ex] \|u\|_{V_A^*\otimes_\e W},\, \|v\|_{V_B^*\otimes_\e W^*}\leq 1}} v\left( (\op{z}_G\otimes \id_W)(u) \right) \\
&=  \sup_{\substack{\\ \dim W\leq L_\leftrightarrow \\[.3ex] \|u\|_{V_A^*\otimes_\e W} \leq 1}} \left\| (\op{z}_G\otimes \id_W)(u) \right\|_{V_B\otimes_\pi W} \\
&= \sup_{\dim W\leq L_\leftrightarrow} \left\| \op{z}_G\otimes \id_W: V_A^*\otimes_\e W \to V_B\otimes_\pi W\right\| ,
\end{align*}
which proves the first upper bound in~\eqref{eq: bounded 2-way bias}. To obtain the other inequalities, we write
\begin{align*}
\left\| \op{z}_G\otimes \id_W: V_A^*\otimes_\e W \to V_B\otimes_\pi W\right\| &\textleq{1} L_\leftrightarrow \left\| \op{z}_G\otimes \id_W: V_A^*\otimes_\pi W \to V_B\otimes_\pi W\right\| \\
&\texteq{2} L_\leftrightarrow \left\| \op{z}_G : V_A^* \to V_B \right\| \\
&\texteq{3} L_\leftrightarrow \left\|z_G\right\|_{V_A \otimes_\e V_B}\, .
\end{align*}
The above steps are easy to justify: 1: we employed the inequality
\bbb
\|\cdot\|_{V_A^*\otimes_\e W} \geq \frac{1}{\dim W}\, \|\cdot\|_{V_A^*\otimes_\pi W} \geq \frac{1}{L_\leftrightarrow}\, \|\cdot\|_{V_A^*\otimes_\pi W}\, ,
\eee
which derives from~\eqref{R} (in turn proven in~\cite[Proposition~21]{ultimate}); 2: follows because the extreme points of the unit ball of $V_A^*\otimes_\pi W$ are product vectors; 3: is an application of~\eqref{injective = operator norm}.
\end{proof}

\begin{remark}
By the same kind of arguments, one can also show that sharing a physical system of bounded dimension does not help to increase the bias by more than a constant factor. We omit the details.
\end{remark}

Finally, let us emphasise that here we have shown an upper bound for the bias of an XOR game with back-and-forth communication. In the work~\cite{Junge18}, the authors studied classical XOR games with both one-way classical communication and one-way quantum communication. It turns out that in that case, the bias of the the games can be exactly expressed in terms of certain norms of the corresponding operator $\op{z}_G:\ell_\infty\to \ell_1$.

\section{Every normed space is \texorpdfstring{$2$}{2}-isomorphic to a base norm space
} \label{app base norm equivalence}

In this appendix we justify our choice of characterising the intrinsic difference between global and local strategies in XOR games by means of the projective/injective ratio as defined by~\eqref{r} instead of~\eqref{r bn}, as discussed in Subsection~\ref{subsec XOR}. This corresponds to letting the optimisation run over all pairs of Banach spaces of fixed dimensions instead of restricting it to the base norm spaces alone, and does not lead to a significant loss of information because of the inequalities~\eqref{r vs r bn same dim} and~\eqref{r vs r bn dim-1}, whose proof we present here. Let us start with a preliminary result.

\begin{lemma} \label{2-isomorphism base norm space lemma}
Every Banach space (possibly infinite-dimensional) is $2$-isomorphic to a base norm space. 
\end{lemma}

\begin{figure}[htbp] 
\begin{center}
\begin{tikzpicture}[scale=1.5]

\coordinate (0) at (0,0);
\coordinate (x) at (0,2);
\coordinate (-x) at (0,-2);
\coordinate (a) at (-1,1);
\coordinate (b) at (1.618,1);
\coordinate (-a) at (1,-1);
\coordinate (-b) at (-1.618,-1);
\coordinate (y) at (2.618,0);
\coordinate (-y) at (-2.618,0);
\coordinate (2a) at (-2,2);
\coordinate (2b) at (3.236,2);
\coordinate (-2a) at (2,-2);
\coordinate (-2b) at (-3.236,-2);
\coordinate (z) at (2.47212, 0.472116);
\coordinate (-z) at (-2.47212, -0.472116);

\draw[fill=gray!30] (2a)--(2b)--(-2a)--(-2b)--(2a);
\draw[fill=gray!90] (a)--(b)--(-a)--(-b)--(a);
\filldraw[black] (0) circle (1pt);
%\draw[->] (0)--(x) node[above]{$x$};
\draw[densely dotted] (x)--(-x);
\draw[densely dotted] (-y)--(y);
\draw[dashdotted] (x)--(z)--(-x)--(-z)--(x);
\draw[black] (0,2) .. controls (0.6,1.9) and (1.2,1.6) .. (1.618,1);
\draw[black] (0,2) .. controls (-0.3,1.99) and (-0.7,1.45) .. (-1,1);
\draw[black] (1.618,1) .. controls (1.709,0.85) .. (1.85,0.5);
\draw[black, densely dashed] (1.85,0.5) -- (1.92,0.304);
\draw[black, densely dashed] (-1,1) .. controls (-1.15,0.75) .. (-1.28,0.5);
\filldraw[black] (1.85,0.5) circle (0.618pt);

\node[] at (0,2.1) {$x$};
\node[] at (0,-2.1) {$-x$};
\node[] at (1.95,0.58) {$y$};
\node[] at (0.7,1.15) {$x^{*}=1/2$};
\node[] at (0.88,2.15) {$x^{*}=1$};
\node[] at (-1.1,1.1) {$a$};
\node[] at (1.718,1.1) {$b$};

\end{tikzpicture}
\end{center}
\caption{\label{fig 2-isomorphic} Geometric intuition behind the proof of Lemma~\ref{r vs r bn lemma}, to which we refer for notation. The plane depicted is that spanned by the vectors $x$ and $y$. From now on, all sets we consider are understood to be intersected with this plane. The curved solid and dashed line represents part of the boundary of the unit ball $B_{X}$ (we set $\|y\|=1$ for simplicity). The points $a$ and $b$ are the intersections of the line $x^{*}=1/2$ with the boundary of $B_{X}$. The associated parallelogram $B$ is depicted in dark grey colour, while $2B$ is in light grey colour. Then, an informal version of the argument is as follows. Since between $x$ and $b$ the boundary of $B_{X}$ lies above the dash-dotted line, convexity implies that it must lie below that same line between $b$ and $y$. This ensures that $y$ belongs to the dash-dotted tilted parallelogram, which is readily verified to fit inside $2B$.}  
\end{figure}
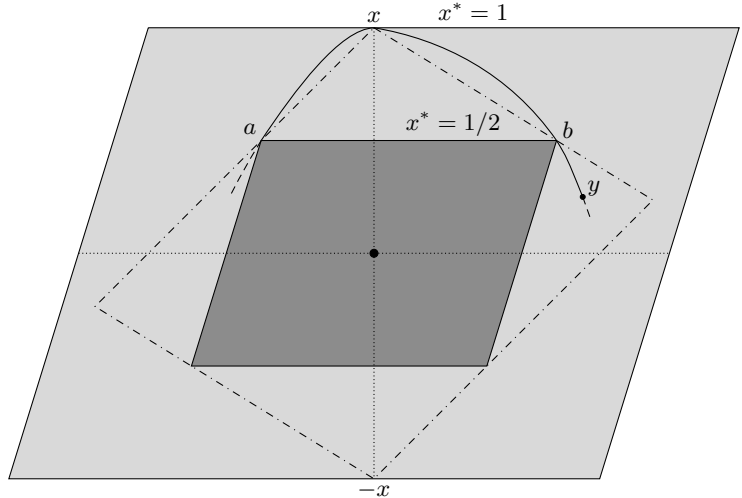

\begin{proof}
Let $X$ be a Banach space. Pick a unit vector $x\in X$ such that $\|x\|=1$, and consider the associated norming functional $x^*\in X^*$, which satisfies $\|x^*\|=1$ and $x^*(x)=1$. Calling $B_X$ the unit ball of $X$, construct the set $F\coloneqq (x^*)^{-1}(1/2)\cap B_X$, and then set $B\coloneqq \cl \conv\left(F \cup (-F)\right)$, where the closure is possibly needed only in the infinite-dimensional case. It is not difficult to verify that $B$ is the unit ball of the base norm space induced on $X$ by the positive cone $\R_+\!\cdot\! F=\{x\in X:\, x^*(x)\geq \|x\|/2\}$ and the unit functional $u\coloneqq 2x^*$~\cite[p.~26]{FOUNDATIONS}. Since $B\subseteq B_X$, it suffices to check that $B_X\subseteq 2B$ to establish the claim. To this end, we pick $y\in B_X$ and we check that $\frac{y}{2}\in B$. We can assume without loss of generality that $x^*(y)\geq 0$, while $|x^*(y)|\leq 1$ holds by construction. We now distinguish two cases.
\begin{itemize}
    \item If $x^*(y)\geq 1/2$, we can write 
    \bbb
    \frac{y}{2} = x^*(y) \frac{y}{2 x^*(y)} \in x^*(y) B \subseteq B\, ,
    \eee
    where we used the fact that $2x^*(y)\geq 1$.
    \item The case where $0\leq x^*(y)<1/2$ is significantly less transparent. Figure~\ref{fig 2-isomorphic} conveys the geometric intuition behind the proof. An analytical argument is as follows. Call $k\coloneqq x^*(y)$, and define the two vectors
    \bbb
        z_\pm \coloneqq \frac{1}{2(1\mp k)}\left( \pm (1\mp 2k) x +y\right) .
    \eee
    
    Observe that $z_+$ lies at the intersection of the segment joining $y$ and $x$ with the plane $x^*=1/2$. Analogously, $z_-$ lies at the intersection of the segment joining $y$ and $-x$ with the plane $x^*=-1/2$. In particular, $z_\pm\in B$. We now try to obtain a multiple of $y$ by taking a convex combination of $z_+$ and $z_-$. Setting
    \bbb
    p(k) \coloneqq \frac{(1+2k)(1-k)}{2(1-2k^2)}\, ,
    \eee
    which satisfies $1/2 \leq p(k) < 1$ for all $0\leq k<1/2$, we can write
    \bbb
    \frac{y}{2(1-2k^2)} = p(k) z_+ + (1-p(k)) z_- \in B\, .
    \eee
    By rescaling the vector on the l.h.s.\ we see that $y/2\in B$.
\end{itemize}
This concludes the proof.
\end{proof}

We are now ready to prove the inequalities~\eqref{r vs r bn same dim} and~\eqref{r vs r bn dim-1} discussed in the main text.

\begin{lemma} \label{r vs r bn lemma}
The functions $r(n,m)$ and $r_{\mathrm{bn}}(n, m)$ defined by~\eqref{r} and~\eqref{r bn} satisfy~\eqref{r vs r bn same dim} and~\eqref{r vs r bn dim-1}:
\begin{align*}
r_{\mathrm{bn}}(n, m) &\leq 4\, r(n, m)\, ,\\
r_{\mathrm{bn}}(n, m) &\leq 2 + r(n-1, m-1)\, ,
\end{align*}
for all integers $n,m\geq 2$.
\end{lemma}

\begin{proof}
Lemma~\ref{2-isomorphism base norm space lemma} proves that for all Banach spaces $X$ there is a base norm Banach space $X'$ such that $\dbm(X,X')\leq 2$, where $\dbm$ is the Banach--Mazur distance~\eqref{Banach-Mazur distance}. We apply this to a pair of finite-dimensional Banach spaces $X,Y$, with $\dim X=n$ and $\dim Y=m$, obtaining two base norm Banach spaces $X'$ and $Y'$ of the same dimension that are $2$-isomorphic to $X$ and $Y$, respectively. We find that
\begin{align*}
r_{\mathrm{bn}}(n,m) &\leq \rho(X',Y') \\
&\leq \dbm(X',X) \rho(X,Y') \\
&\leq \dbm(X',X) \dbm(Y',Y) \rho(X,Y) \\
&\leq 4\rho(X,Y)\, ,
\end{align*}
where we used~\eqref{eq: rho heredity d} twice, once for each of the two arguments of $\rho$ (this is possible as $\rho$ is symmetric, see~\eqref{eq: rho dual and exchange}). Taking the infimum over all pairs $X,Y$ yields~\eqref{r vs r bn same dim}.

We now move on to proving the second inequality~\eqref{r vs r bn dim-1}. The main idea of the argument is to construct, given a pair of Banach spaces $X,Y$ of dimensions $n-1,m-1$, another pair of base norm spaces $X',Y'$ of dimensions $n,m$ such that $\rho(X',Y')\approx \rho(X,Y)$. This can be done by setting $X'\coloneqq X\oplus_\infty \R$, where $\|(x,a)\|_{X'}= \max\{\|x\|_X,|a|\}$ for all $x\in X$ and $a\in \R$, and analogously for $Y'$. It is not difficult to check that $\|\cdot\|_{X'}$ is in fact the base norm induced by the cone $C\coloneqq \{(x,a): a\geq \|x\|_X\}$ and the order unit $u_{X'}\in (X')^*$ defined by $u_{X'}(x,a)=a$ for all $x\in X$ and $a\in \R$. Thus, $X'$ and $Y'$ are base norm spaces. Incidentally, this is a systematic way of associating `centrally symmetric' GPTs to Banach spaces, see~\cite[Section~6.1]{ultimate}. We now proceed to show that 
\bb
\rho(X',Y')\leq 2 + \rho(X,Y)\, ,
\label{rho vs rho'}
\ee
using a similar technique to that employed in the proof of~\cite[Proposition~26]{ultimate}. Take 
\bbb
z' = \begin{pmatrix} z & s \\ t & a \end{pmatrix}\in X'\otimes Y'\, ,
\eee
where $z\in X\otimes Y$, $s\in X$, $t\in Y$, and $a\in \R$. Using the fact that a unit functional $\varphi\in B_{(X')^*}$ acts as $\varphi(x,a) = p x^*(x) \pm (1-p)a$, for some $x^* \in B_{X^*}$ and $p\in [0,1]$, it is not difficult to show that
\bb
\|z'\|_{X'\otimes_\e Y'} = \max\left\{ \|z\|_{X\otimes_\e Y}, \|s\|, \|t\|, |a| \right\} .
\label{r vs r bn: epsilon norm z'}
\ee
We now give an upper estimate of the corresponding projective norm. Taking vectors $x_i\in X$ and $y_i\in Y$ such that $z=\sum_i x_i\otimes y_i$ and $\|z\|_{X\otimes_\pi Y} = \sum_i \|x_i\| \|y_i\|$, and for an arbitrary $p\in [0,1]$, we consider the decomposition
\bbb
z' = \begin{pmatrix} 0 & s \\ 0 & pa \end{pmatrix} + \begin{pmatrix} 0 & 0 \\ t & (1-p)a \end{pmatrix} + \begin{pmatrix} z & 0 \\ 0 & 0 \end{pmatrix} = (s,pa)\otimes (0,1) + (0,1)\otimes (t,(1-p)a) + \sum_i (x_i,0)\otimes (y_i,0)\, ,
\eee
which yields the estimate
\bb
\begin{aligned}
\|z'\|_{X'\otimes_\pi Y'} &\leq \min_{p\in [0,1]} \left\{ \max\{\|s\|,p|a|\} + \max\{\|t\|,(1-p)|a|\} +\sum_i \|x_i\|\|y_i\|\right\} \\
&= \max\left\{\|s\|+\|t\|,|a|\right\} + \sum_i \|x_i\| \|y_i\| \\
&= \max\left\{\|s\|+\|t\|,|a|\right\} + \|z\|_{X\otimes_\pi Y} \\[1ex]
&\leq \max\left\{\|s\|+\|t\|,|a|\right\} + \rho(X,Y) \|z\|_{X\otimes_\e Y} \\[1ex]
&\leq \left( 2 + \rho(X,Y)\right) \|z'\|_{X'\otimes_\e Y'}\, .
\end{aligned}
\ee
Optimising over all $z'\in X'\otimes Y'$ and using~\eqref{rho} gives the estimate in~\eqref{rho vs rho'}. We can now write
\begin{align*}
    r_{\mathrm{bn}}(n,m) &\leq \inf_{\substack{\dim X = n-1 \\ \dim Y = m-1}} \rho(X',Y') \\
    &\leq \inf_{\substack{\dim X = n-1 \\ \dim Y = m-1}} \left\{ 2 + \rho(X,Y) \right\} \\
    &= 2 + r(n-1,m-1)\, .
\end{align*}
This concludes the proof.
\end{proof}

\section{Functional analytic tools}

\label{app functional analysis}

%This need not be a separate appendix, some part (all?) of this could be also included in the main text.

\subsection{The \texorpdfstring{$\ell$}{ell}-norm and the \texorpdfstring{$MM^*$}{MM*}-estimate}

Let $X$ be a real  Banach space. Given a linear map $T:\ell_2^n\rightarrow X$, the {\bfseries $\ell$-norm} of $T$ is defined as 
\begin{align}\label{ell-norm}
\ell_X(T)= \E\left\|\sum_{i=1}^ng_i T|e_i\rangle\right\|_X,
\end{align}
%\begin{align}\label{ell-norm}
%\ell(T)= \left(\E\left\|\sum_{i=1}^ng_i T|e_i\rangle\right\|_X^2\right)^{\frac{1}{2}},
%\end{align}
where $(|e_i\rangle)_{i=1}^n$ is an orthormal basis of $\mathbb R^n$ and $(g_i)_{i=1}^n$ is a sequence of independent $N(0,1)$ Gaussian random variables. We point out that several authors prefer to define $\ell$-norms via the second moment, i.e.\ $\ell_X(T)=  \left( \E\left\|\sum_{i=1}^ng_i T|e_i\rangle\right\|_X^2 \right)^{1/2}$. However both definitions give equivalent norms in view of Theorem~\ref{Khinchin-Kahane}. Also, note that the invariance of the Gaussian measure under unitary transformations implies that $\ell_X(T)=\ell_X(T\circ U)$ for every unitary $U:\ell_2^n\rightarrow \ell_2^n$ and, in particular,~\eqref{ell-norm} does not depend on the choice of orthonormal basis.

%Trace duality allows us to define the dual norm of $\ell$. Indeed, given any Banach space $X$ and any linear map $S:X\rightarrow \ell_2^n$, the \emph{$\ell^*$-norm} of $S$ is defined as 
%\begin{align}\label{ell^*-norm}
%\ell^*(S)= \sup \{|\tr(S\circ T)|: \ell(T:\ell_2^n\rightarrow X)\leq 1\}.
%\end{align}

The following theorem will be crucial for us.
\begin{theorem}[$MM^*$-estimate]\label{ell-K}
Let $X$ be an $n$-dimensional Banach space. Then there exists an isomorphism $T:\ell_2^n\rightarrow X$ such that
\begin{align*}
\ell_X(T)\ell_{X^*}((T^{-1})^*)\leq Cn \log n.
\end{align*}
%\item[1. ] There exists an isomorphism $S:\ell_2^n\rightarrow X$ such that $$\|S\| \cdot \|S^{-1}\|=d_X\leq \sqrt{n}.$$
%\item[2. ] There exists an isomorphism $T:\ell_2^n\rightarrow X$ such that
%\begin{align*}
%\ell(T)\ell^*(T^{-1})=n.
%\end{align*}
%In addition, 
%\begin{align*}
%\ell(T)\ell((T^{-1})^*)\leq Kn \log (1+d_X),
%\end{align*}where $K$ is a universal constant and $d_X=d(X, \ell_2^n)$ is the Banach--Mazur distance between $X$ and $\ell_2^n$.
%\end{enumerate}
\end{theorem}

%The first part of the theorem is a re-statement of John ellipsoid theorem (see~\cite[Propostion~3.8]{Pisier99} for an even stronger result). \textbf{Any better reference?}

That statement is a direct consequence of Lewis' theorem (\cite[Theorem~3.1]{Pisier99}) and a well known estimate on the so-called $K$-convexity constant of a Banach space. The reader can find a detailed proof of Theorem~\ref{ell-K} in~\cite[Theorem~3.11]{Pisier99} or~\cite[Theorem~7.10]{ABMB}.

\subsection{Some Gaussian inequalities}

We will make use of Khintchine--Kahane inequalities (see for instance~\cite[Corollary~4.9]{Pisier99} or, for optimal constants, 
\cite[Corollary~3]{LO99}).
\begin{theorem}[Khintchine--Kahane inequalities]\label{Khinchin-Kahane}
For every $1< p<\infty$  
%actually $0\leq p<\infty$, $p=0$ corresponding to the `geometric mean'
there exists a universal constant $C_p>0$ such that for every Banach space $X$ and every sequence of elements $(x_i)_{i=1}^n\subset X$ we have 
\begin{align*}
\E\left\|\sum_{i=1}^n g_ix_i\right\|_X\leq \left(\E\left\|\sum_{i=1}^n g_i x_i\right\|_X^p\right)^{\frac{1}{p}}\leq C_p \E\left\|\sum_{i=1}^n g_i x_i\right\|_X.
\end{align*}
\end{theorem}

We will also make use of Chevet's inequality (\cite[Theorem~43.1]{TOMCZAK}).
\begin{theorem}[Chevet's inequality]\label{Chevet}
Let $X$ and $Y$ be real Banach spaces. Define the Gaussian random tensor $z=\sum_{i=1}^m \sum_{j=1}^n g_{ij}x_i\otimes y_j\in X\otimes Y$, where $(g_{ij})$ are independent $N(0,1)$ Gaussian random variables, and $(x_i)_{i=1}^m \subset X$, $(y_j)_{j=1}^n\subset Y$ are sequences of elements. Then,
\begin{align*}
\E \left\|z\right\|_{X \otimes_{\e} Y}\leq \sup_{x^*\in B_{X^*}}\left(\sum_{i=1}^m |x^*(x_i)|^2\right)^{\frac{1}{2}} \E\left\|\sum_{j=1}^n g_j y_j \right\|_{Y}+ \sup_{y^*\in B_{Y^*}}\left(\sum_{j=1}^n |y^*(y_j)|^2\right)^{\frac{1}{2}}  \E\left\|\sum_{i=1}^m g_i x_i \right\|_{X}, 
\end{align*}
where $(g_{i})_i$ is a sequence % and $(g_{j})_{j=1}^n$ denote sequences 
of independent $N(0,1)$ Gaussian random variables.
\end{theorem}

Note that, given a Banach space $Z$ and $(z_i)_{i=1}^n\subset Z$, we have $$\sup_{z^*\in B_{Z^*}}\left(\sum_{i=1}^n |z^*(z_i)|^2\right)^{\frac{1}{2}}=\|T:\ell_2^n\rightarrow Z\|.$$Here $T$ is the linear map defined by $T|e_i\rangle =z_i$ for every $i=1,\cdots, n$, where $(|e_i\rangle)_i$ is an orthonormal basis of $\ell_2^n$.

Here is a typical application of Chevet's inequality. Fix integers $m$, $n$, and consider $(x_i)_{1 \leq i \leq m^2}$ and $(y_j)_{1 \leq j \leq n^2}$ orthonormal bases of $\Herm{m}$ and $\Herm{n}$ respectively, with respect to the Hilbert--Schmidt inner product. We form the random tensor
\begin{equation} \label{eq:standard-gaussian-schatten} z = \sum_{i=1}^{m^2} \sum_{j=1}^{n^2} g_{ij} x_i \otimes y_j, 
\end{equation}
where $(g_{ij})$ are independent $N(0,1)$ Gaussian random variables.

\begin{corollary}
\label{chevet-schatten}
Let $z$ be defined as in~\eqref{eq:standard-gaussian-schatten}. Remember that we denote by $S_{p}^{n,\mathrm{sa}}$ the space of $n\times n$ Hermitian matrices equipped with the Schatten norm $\|\cdot\|_p$. Then
\[ \E \|z\|_{\Tm \otimes_{\e} \Tn} \leq C \sqrt{mn} \max\{m,n\} ,\]
\[ \E \|z\|_{S_{\iy}^{m,\mathrm{sa}} \otimes_{\e} S_{\iy}^{n,\mathrm{sa}}} \leq C \max\{\sqrt{m},\sqrt{n}\}. \]
\end{corollary}

\begin{proof}
In both cases we apply Theorem~\ref{Chevet} and need to estimate all quantities appearing on the r.h.s. The random matrix $G_m = \sum_{j=1}^m g_j x_j$ is distributed according to the Gaussian Unitary Ensemble. It is well known (see for example~\cite[Proposition 6.24]{ABMB}) that as $m$ tends to infinity, 
\[ \E \|G_m\|_{\iy} \sim 2 \sqrt{m}, \ \ \ \E \|G_m\|_{1} \leq m^{3/2} .\]
On the other hand, it is also well known (and easy to check) that 
\[  ||\id:S_2^{m,\mathrm{sa}}\rightarrow S_\infty^{m,\mathrm{sa}}||\leq 1, \ \ \ ||\id:S_2^{m,\mathrm{sa}}\rightarrow S_1^{m,\mathrm{sa}}||\leq \sqrt{m}.\]
Since $S_2^{m,\mathrm{sa}}$ is isometric to $\ell_2^{m^2}$, we can apply Theorem~\ref{Chevet} to conclude that
\[ \E \|z\|_{\Tm \otimes_{\e} \Tn} \leq n^{\frac{3}{2}}\sqrt{m}+ m^{\frac{3}{2}}\sqrt{n} \leq 2\sqrt{mn} \max\{m,n\},\]
\[ \E \|z\|_{S_{\iy}^{m,\mathrm{sa}} \otimes_{\e} S_{\iy}^{n,\mathrm{sa}}} \leq 2 \sqrt{m}+ 2\sqrt{n} \leq 4 \max\{m,n\}; \]
hence the result follows.
\end{proof}

Finally, we will use the following lemma, whose proof is elementary. %\textbf{I did not find a reference for it. The proof is easy. But I think that adding a proof for this lemma is too much.}

\begin{lemma}[Contraction principle]\label{Gaussian-Lemma}
Let $(\alpha_i)_{i=1}^n$ and $(\beta_i)_{i=1}^n$ be two sequences of numbers with $0\leq \alpha_i\leq \beta_i$ for every $i$. Let $(g_i)_{i=1}^n$ be a sequence of independent $N(0,1)$ Gaussian random variables. Then, for every Banach space $X$ and every $x_1,\cdots, x_n\in X$, we have \begin{align*}
\E\left\|\sum_{i=1}^n\alpha_i g_ix_i\right\|_X\leq \E\left\|\sum_{i=1}^n \beta_i g_ix_i\right\|_X.
  \end{align*}
\end{lemma}

\subsection{Dvoretkzy--Milman theorem}

We also need Milman's version of Dvoretzky theorem (see e.g.~\cite[Theorem 7.19]{ABMB}). Let $\|\cdot\|_X$ be a norm on $\R^n$ and consider the space $X = (\R^n,\|\cdot\|_X)$. The {\bfseries Dvoretzky dimension} of $X$ is defined as 
\[ k_*(X) = \left(\frac{\ell_X(\id)}{\| \id : \ell_2^n \to X \|}\right)^2. \]

\begin{theorem}[Dvoretzky--Milman theorem] \label{theorem: dvoretzky}
Consider a normed space $X=(\R^n,\|\cdot\|_X)$ and let $E \subseteq \R^n$ be a random subspace of dimension $k \leq c\, k_*(X)$. Then, with large probability,
\bbb
\frac{\ell_X(\id)}{2\sqrt{n}} |x| \leq \|x\|_X \leq \frac{2\ell_X(\id)}{\sqrt{n}} |x|
\eee
for every $x \in E$, where $|\cdot|$ is the standard Euclidean norm on $\R^n$. In particular, the space $X \cap E$ is $4$-Euclidean.
\end{theorem}

\begin{remark}
In Theorem~\ref{theorem: dvoretzky} it is understood that $E$ is distrbuted according to the Haar measure on the Grassmann manifold (see e.g.~\cite{ABMB}). The expression `with large probability' means that the probability of failure tends to zero exponentially fast as $n$ tends to infinity; we need only to know that the intersection of two such events of large probability is nonempty.
\end{remark}

\subsection{John ellipsoid}

The following theorem is a classical result about convex bodies (see~\cite{Ball97} for a modern proof).

\begin{theorem}[John's theorem] \label{theorem:john}
For every $n$-dimensional normed space $X$ with unit ball $B_X$, there is a unique ellipsoid $\mathcal{E}$ of maximal volume under the constraint $\mathcal{E} \subseteq B_X$. The ellipsoid $\mathcal{E}$ is called the John ellipsoid of $X$ and satisfies $B_X \subseteq \sqrt{n} \mathcal{E}$. Consequently, we have $\dbm(X,\ell_2^n) \leq \sqrt{n}$.
\end{theorem}

We also use a variant of John's theorem. It can be for example deduced from \cite[Corollary 3.9]{Pisier99}.

\begin{proposition} \label{proposition:john-factorization}
Let $X$ be a finite-dimensional normed space with $\dim (X) \geq n$. Then there exist maps $u : \ell_2^n \to X$ and $v:X \to \ell_2^n$ such that $vu = \id_{\ell_2^n}$, $\|u\|=1$ and $\|v\| \leq \sqrt{n}$.
\end{proposition}

We will rely on a technical result which guarantees that certain normed spaces contains large-dimensional cubes; in that formulation it is due to Vershynin~\cite{Vershynin01} (improving on Rudelson~\cite{Rudelson95}).

\begin{theorem}[Theorem 6.2 in~\cite{Vershynin01}] \label{theorem: vershynin}
Let $X=(\R^n,\|\cdot\|_X)$ be a $n$-dimensional normed space whose John ellipsoid is $B_2^n$. Let $P$ be an orthogonal projection and $k = \mathrm{rank} (P)$. 
Then there are $m \geq c k/\sqrt{n}$ contact points $(x_j)_{1 \leq j \leq m}$ such that
\begin{equation} \label{eq:vershynin} \max_{1 \leq j \leq m} | \scalar{x}{x_j} | \leq \|x\|_X \leq C \sqrt{\frac{n}{k}}\ \ell_X(P) \max_{1 \leq j \leq m} | \scalar{x}{x_j} |
\end{equation}
for every $x \in \mathspan \{ Px_j \st 1 \leq j \leq m \}$. In particular, the space $X$ contains a subspace which is $R$-isomorphic to $\ell_{\iy}^m$ for $R=C \ell_X(P) \sqrt{n/k}$.
\end{theorem}

%\subsection{Proportional Dvoretzky--Rogers factorisation}

%\tcr{Possibly not any more needed?}

%Another result which we use is the following variation on the Dvoretzky--Rogers lemma, which is due to Bourgain and Szarek~\cite{BourgainSzarek88}.

%\begin{theorem}[Proportional Dvoretzky--Rogers factorisation] \label{theorem:bourgainszarek}
%There are constants $\delta>0$, $D$ such that every $n$-dimensional normed space
%contains $m=\delta n$ vectors $\{x_i\}_{i=1,\ldots, m}$ such that, for any set of scalars $\{t_i\}_{i=1,\ldots, m}$, one has that
%\[  \max_{1 \leq i \leq m} |t_i|  \leq \left\| \sum_{i=1}^m t_i x_i \right\| \leq D \left( \sum_{i=1}^m t_i^2 \right)^{1/2}. \]
%\end{theorem}

%\begin{remark}
%In Theorem~\ref{theorem:bourgainszarek}, we can actually require $\delta$ to be arbitrarily close to $1$, with $D$ depending on $\delta$. However, we will not need this refinement here. The present formulation is substantially easier to prove, see~\cite[Remark~6]{BourgainSzarek88}.
%\end{remark}

\subsection{\texorpdfstring{$\boldsymbol{p}$}{p}-summing norms}
\label{app p-summing}

Let $u : X \to Y$ be a linear map between finite-dimensional normed spaces. Fix $p \in [1,\iy)$; we only need $p=1$ and $p=2$ in the present paper. For an integer $N$, we define a quantity $\pi_p^{(N)}(u)$ to be the smallest constant $K$ such that, for every $N$ vectors $x_1,\cdots,x_N \in X$, we have
\begin{equation} \label{def p-summing} \left( \sum_{k=1}^N \|u(x_k)\|^p \right)^{1/p} \leq K \sup_{\phi \in B_{Y^*}}  \left( \sum_{k=1}^N |\phi(x_k)|^p \right)^{1/p}. \end{equation} 
The quantity $\pi_p(u) = \sup \{ \pi_p^{(N)}(u) \st N \geq 1\}$ is called the {\bfseries $\boldsymbol{p}$-summing norm} of the operator $u$.

\begin{proposition} \label{proposition p-summing}
Consider finite-dimensional normed spaces $X, Y$, and a linear operator $u : X \to Y$. We have
\begin{enumerate}
    \item $\pi_1^{(N)}(u) \geq \pi_2^{(N)}(u)$,
    \item $\pi_2^{(\dim X)}(u) \geq \frac{1}{\sqrt{2}} \pi_2(u)$,
    \item $\pi_2(\id_X) = \sqrt{\dim X}$.
\end{enumerate}
\end{proposition}

A general reference about $p$-summing norms (with detailed bibliography) is \cite{TOMCZAK}; parts 1--3 of Proposition \ref{proposition p-summing} appear there respectively as Proposition 9.6, Theorem 18.4 and Proposition 9.11. 

We also need a specific result about the $1$-summing norm of the identity map on $\ell_1^n$, which appears as \cite[Theorem 2(4)]{Gordon69} : we have
%\begin{proposition} %\label{proposition:1-summing-ell1}
%We have 
\begin{equation} \label{eq:1-summing-ell1} \pi_1(\id_{\ell_1^n}) = \frac{n}{ \E \left| \sum_{i=1}^n \e_i \right| }, \end{equation}
where $(\e_i)$ is a sequence of independent random variables with $\P(\e_i=1)=\P(\e_1=-1) = \frac 12$. For a more transparent derivation, one may use the fact that $\ell_1^n$ has enough symmetries in the sense of \cite[\S 16]{TOMCZAK}, which implies that $\pi_1(\id_{\ell_1^n}) = n/\f (\ell_1^n,\ell_\infty)$ (see e.g.\ \cite{KonigTomczak90}). In turn, the quantity $\f (X,\ell_\infty)$ (defined in \eqref{factorisation constant}  and also referred to as the \textbf{projection constant} of a normed space $X$, see \cite[\S 32]{TOMCZAK}) can be calculated directly when $X=\ell_1^n$ and equals $\E |\e_1 + \cdots + \e_n|$; an early reference for the last result is \cite[Theorem 3]{Grunbaum60}.

%\end{proposition}

%\begin{proof}
%We use a result from \cite{KonigTomczak90}: since $\ell_1^n$ has enough symmetries in the sense of \cite[\S 16]{TOMCZAK}, we have $\pi_1(\id_{\ell_1^n}) = n/\lambda(\ell_1^n)$, where $\lambda(X)$ is the \textbf{projection constant} of a normed space $X$, as defined in \cite[\S 32]{TOMCZAK}. To complete the proof one needs to know the fact that $\lambda(\ell_1^n) = \E |\e_1 + \cdots + \e_n|$; this result appears in \cite[Theorem 3]{Grunbaum60}.
%\end{proof}

\subsection{Non-commutative Grothendieck inequality}

Let us recall Grothendieck's inequality for bilinear forms on $C^*$-algebras. Here, we state~\cite[Theorem~1.1]{Haagerup85}, which improved the original proof in~\cite{Pisier78}.

\begin{theorem}[Grothendieck's inequality for $C^*$-algebras] \label{Grothendieck C* algebras thm}
Let $V:\A\times \B\rightarrow \mathbf{C}$ be a bilinear form on a pair of $C^*$-algebras $\A$ and $\B$. Then, there exist two states $\varphi_1$ and $\varphi_2$ on $\A$ and two states $\psi_1$ and $\psi_2$ on $\B$ such that
\begin{equation}
|V(x,y)|\leq \|V\|\left(\varphi_1(x^*x)+\varphi_2(xx^*)\right)^{\frac{1}{2}}\left(\psi_1(y^*y)+\psi_2(yy^*)\right)^{\frac{1}{2}} \qquad \forall\ x\in \A, \, y\in \B ,
\end{equation}
where
\begin{equation}
    \|V\|\coloneqq \sup_{\substack{\\ x\in \A,\, y\in \B\\[.3ex] \|x\|, \|y\|\leq 1 }} |V(x,y)|
\end{equation}
is the norm of $V$.
\end{theorem}

By applying the previous theorem to the particular case $\A=S_{\iy}^n$ (the $C^*$-algebra of $n \times n$ complex matrices endowed with the operator norm) and $\B=S_{\iy}^m$, we deduce the following corollary.

\begin{corollary}\label{cor state}
Let $z\in \Tn\otimes_\e \Tm$ be a tensor, and let $\op{z}:S_{\iy}^{n, \mathrm{sa}}\to \Tm$ be the linear map associated with it according to~\eqref{tensor-operator}. Then, there exists a state $\varphi$ on $S_{\iy}^{n}$ such that
\begin{equation}
    \|\op{z}(x)\|_{\Tm}\leq 2\sqrt{2}\, \|z\|_{\Tn\otimes_\varepsilon \Tm}\big(\varphi(x^2)\big)^{\frac{1}{2}}\qquad  \forall\ x\in S_\iy^{n,\mathrm{sa}}\, .
    \label{eq:quantum-Grothendieck}
\end{equation}
\end{corollary}

\begin{proof}
We start by remarking that since $\Tk$ can be thought of as a (real) subspace of the (complex) Banach space of all $k\times k$ complex matrices endowed with the trace norm, denoted by $S_1^k$, we can consider $z$ also as a tensor in $S_1^n\otimes_\e S_1^m$. According to~\cite[Claim~4.7]{RV13}, we have that
\begin{equation*}
    \|z\|_{S_1^n\otimes_{\e} S_1^m}\leq \sqrt{2}\, \|z\|_{\Tn \otimes_\e \Tm}.
\end{equation*}
Indeed, to see this just notice that~\cite[Definition~4.3]{RV13} and~\cite[Definition~4.6]{RV13} correspond to $\|z\|_{\Tn\otimes_\e \Tm}$ and $\|z\|_{S_1^n\otimes_\e S_1^m}$, respectively. Now we consider the bilinear form $V_z: S_\iy^n\times S_\iy^m\to\mathbf{C}$ defined by $V_z(x,y)\coloneqq \Tr\left[(x\otimes y)z\right]$, whose norm can be verified to coincide with the injective norm of the tensor $z$, i.e.\ 
\begin{equation*}
    \|V_z\| = \|z\|_{S_1^n\otimes_{\e} S_1^m} .
\end{equation*}
Applying Theorem~\ref{Grothendieck C* algebras thm} to $V_z$ then yields
\begin{align*}
    \left|\Tr[\op{z}(x) y] \right| &= \left|\Tr[(x\otimes y) z]\right| \\
    &\leq \|z\|_{S_1^n\otimes_{\e} S_1^m} \left(\varphi_1(x^*x)+\varphi_2(xx^*)\right)^{\frac{1}{2}} \left(\psi_1(y^*y)+\psi_2(yy^*)\right)^{\frac{1}{2}} \\
    &\leq \sqrt{2}\, \|z\|_{\Tn \otimes_{\e} \Tm} \left(\varphi_1(x^*x)+\varphi_2(xx^*)\right)^{\frac{1}{2}} \left(\psi_1(y^*y)+\psi_2(yy^*)\right)^{\frac{1}{2}} .
\end{align*}
Taking the supremum over all $y\in S_{\infty}^{m,\mathrm{sa}}$ such that $\|y\|_{\iy}\leq 1$, using the fact that $\psi(y^*y)\leq 1$ and $\psi(yy^*)\leq 1$ for all such $y$ and for all states $\psi$ on $S_{\infty}^m$, and finally defining $\varphi\coloneqq (\varphi_1+\varphi_2)/2$, we obtain precisely~\eqref{eq:quantum-Grothendieck}.
%According to this, one immediately deduces that $\|B_z\|\leq \sqrt{2}\|z\|_{\Tn\otimes_{\varepsilon} \Tm}$, where $B_z:M_n\times M_m\rightarrow \mathbb C$ denotes the bilinear form associated to $z$, which coincides with $\|z\|_{S_1^n\otimes_{\varepsilon} S_1^m}$. Then, we deduce
%\begin{align*}\|T_z(x)\|_{\Tm}&=\sup_{\substack{y\in M_n^{\mathrm{sa}}\\\|y\|_{M_n}\leq 1}}|\langle T_z(x), y\rangle|=\sup_{\substack{y\in M_n^{\mathrm{sa}}\\\|y\|_{M_n}\leq 1}}|B_z(x,y)|\\&\leq \sup_{\substack{y\in M_n^{\mathrm{sa}}\\\|y\|_{M_n}\leq 1}}\|B_z\|\Big(\varphi_1(x^2)+\varphi_2(x^2)\Big)^{\frac{1}{2}}\Big(\psi_1(y^2)+\psi_2(y^2)\Big)^{\frac{1}{2}}\\&\leq 2\sqrt{2}\|z\|_{\Tn\otimes_{\varepsilon} \Tm}\big(\varphi(x^2)\big)^{\frac{1}{2}},
%\end{align*}where the first inequality follows from the previous theorem and in the last inequality we have defined $\varphi=(\varphi_1+\varphi_2)/2$.
\end{proof}

\end{document}